\documentclass[aip,amsfonts,amssymb,amsmath,reprint,showkeys,onecolumn,nofootinbib]{revtex4-1}
\usepackage[english]{babel}
\usepackage[utf8]{inputenc}
\usepackage{amsthm}
\usepackage{mathtools}
\usepackage{physics}
\usepackage{xcolor}
\usepackage{wrapfig}
\usepackage{subcaption}
\usepackage{graphicx}
\usepackage{mathtools}

\usepackage{graphicx,epsfig}
\usepackage[left=23mm,right=13mm,top=35mm,columnsep=15pt]{geometry} 
\usepackage{adjustbox}
\usepackage{placeins}
\usepackage[T1]{fontenc}
\usepackage{csquotes}
\usepackage{stackengine,scalerel}

\usepackage[pdftex, pdftitle={Article}, pdfauthor={Author}]{hyperref} 
\newcommand{\probP}{\text{I\kern-0.15em P\ }}
\newcommand{\fC}{f_{\mathcal{C}}}

\newcommand{\modurc}{URC{ }}
\newcommand{\modrgm}{RGM{ }}
\newcommand{\modegm}{EGM{ }}

\begin{document}
\title{Clusters in randomly-coloured spatial networks}

\author{Silvia Rognone}
\affiliation{School of Mathematical Sciences, Queen Mary University of
London, Mile End Road E1 4NS London, UK}
\author{Vincenzo Nicosia}
\affiliation{School of Mathematical Sciences, Queen Mary University of
London, Mile End Road E1 4NS London, UK}


\begin{abstract} The behaviour and functioning of a variety of complex
physical and biological systems depend on the spatial organisation of
their constituent units, and on the presence and formation of clusters
of functionally similar or related individuals. Here we study the
properties of clusters in spatially-embedded networks where nodes are
coloured according to a given colouring process. This characterisation
will allow us to use spatial networks with uniformly-coloured nodes as a
null-model against which the importance, relevance, and significance of
clusters of related units in a given real-world system can be assessed.
We show that even a uniform and uncorrelated random colouring process
can generate coloured clusters of substantial size and interesting
shapes, which can be distinguished by using some simple dynamical
measures, like the average time needed for a random walk to escape from
the cluster. We provide a mean-field approach to study the properties of
those clusters in large two-dimensional lattices, and we show that the
analytical treatment agrees very well with the numerical results.
\end{abstract}

\keywords{spatial networks, spatial patterns, null-models, coloured
networks, random walks, hitting time, }

\maketitle

\section{Introduction}

The characterisation of spatial patterns, and the quantification of their
significance, has pivotal importance in a variety of research and application
fields, from urban studies to biology and
ecology~\cite{Papadimitriou2020,Barthelemy2022}.  Indeed, the functioning of
many physical and biological systems is intimately connected with their spatial
organisation, and the emergence of certain characteristics and behaviours is
almost inevitably underpinned or fostered by the presence (or development) of
specific spatial patterns~\cite{Traveset2012,Louf2014,Bassel2014}.  One example
is the emergence of social, economic, and ethnic segregation in urban
areas~\cite{Batty2005,Batty2013,Barthelemy2017}. The spatial arrangement of
classes or categories throughout a city is normally far from uniform, and
individuals belonging to the same social, economic, or ethnic group are very
often found clustered in specific areas. As a result of these biases, modern
cities normally look pretty fragmented or segregated at several meaningful
scales, a condition that is often connected to social deprivation, uneven
access to resources and services, and heterogeneous distribution of certain
kinds of crimes ~\cite{Bassolas2021,Bassolas2021bis}. Another interesting
example is the emergence of clusters in the development of mutants in
biological systems, as in the case of cancerous masses, where the distribution,
shape, and spatial organisation of those masses has been consistently
associated with the aggressiveness of the tumour ~\cite{Haughey2023}. 

The presence of spatially-segregated clusters of units belonging to a
certain class is normally seen as an interesting property of spatial
complex systems \textit{per-se}, under the assumption that a cluster is
something that does not emerge naturally in a random, unorganised
system. However, this hypothesis is seldom (if ever) tested against a
suitable null-model, in order to associate a significance to the
observed spatial fragmentation of a system under study. So we would
generally accept that a city like London is segregated with respect to
ethnicity, as we observe a quite uneven distribution of
spatially-segregated ethnic groups across its metropolitan area, but
there is no agreement about whether the situation in London is any
better, or worse, or even comparable to the ethnic segregation observed,
say, in Los Angeles or in San Francisco. This problem has been tackled
in a series of recent works on the subject, which have striven to
improve the objectivity of comparisons of structural indicators across
different spatial systems~\cite{Sousa2022,Bassolas2021}.

The aim of this paper is to characterise the shape, size, and geometric
properties of the clusters of units emerging in spatial networks with
coloured nodes. Because of their emergence from a similar random
stochastic process, these clusters can be related to percolation
clusters, and fall into a larger category of objects that are normally
studied in different areas of mathematics and physics, such as
polyominoes and lattice animals~\cite{Golomb1994, Whittington1990,
Xu2014}. The rationale behind our study is that the significance of an
observed spatial pattern can only be assessed with respect to a sound
null-model, and uniform random colourings provide a first-order
null-model for that purpose. Our results show that relatively large
connected clusters of nodes of a single colour are not at all that rare,
even in the case of random uniform colouring processes. This means that
the sheer size of a cluster might not be a good proxy for its
significance, unless a proper comparison with a uniform random colouring
process is carried out. We also show that the typical clusters emerging
in random colourings are in general tree-like and very elongated, thus
allowing us to employ the mean time needed for a random walk to exit a
cluster as a proxy of its randomness. We also provide a mean-field
theory that allows us to predict the expected size of clusters as the
size of the underlying lattice increases. 

\section{Clusters in coloured lattices} 

\subsection{Notation}

We call $\mathcal{G}$ a graph, or a network, defined as a pair
($\mathcal{N}$, $\mathcal{E}$) of the set of nodes $\mathcal{N}$ and the
set $\mathcal{E}$ of pairs of nodes $(i,j)$, where each of these pairs
represents an edge, i.e. a relationship, or a link, between the nodes
$i$ and $j$~\cite{Latora2017}. 
For the purpose of this study, we only take into consideration spatial
graphs~\cite{Barthelemy2022}, whose nodes are associated with specific
points in a given space and edges represent the connections or
relationships among those locations, typically in terms of spatial
proximity, adjacency, connectivity or transfer cost between the two
corresponding points. We start by considering infinite two-dimensional
square lattices, that are infinite graphs where each node $i$ is placed
at one of the intersections $(x_i, y_i)$ of the grid of integer
coordinates (i.e., $x_i, y_i$ are integer numbers, for all values of
$i$), and is connected to the four immediately adjacent nodes placed at
a distance $1$ from it, i.e., the nodes at coordinates $(x_i-1, y_i)$,
$(x_i+1, y_i)$, $(x_i, y_i-1)$, and $(x_i, y_i + 1)$.
Square lattices are widely used as a convenient model for studying
spatial phenomena~\cite{Kawano2019, Grujic2012}, and are largely studied
also in the case of pattern analysis~\cite{Meakin1986, Dill1995}. The
main reason behind this choice is that, although they are a simple model
of real-world spatial systems, square lattices are one of the most
straightforward discretizations of the plane. 

Since our main motivation is to provide a suitable and sound null-model
for patterns observed in real-world spatial networks, we will consider
in the following two concrete finite lattice geometries, namely grids
and tori.  A grid is a square lattice that has a finite size and a
finite boundary. Nodes that are not in the boundary are connected to
four neighbours, while nodes on the boundary will lack one or two
neighbours (depending on whether they are placed on the side or on the
corner of the lattice). As a result, not all nodes in a grid are
structurally indistinguishable, and this fact has an impact on the
estimation of the average properties of the system, especially when the
size of the lattice is small. On the other hand, a torus is still a
lattice with finite size, but it has periodic boundary conditions,
effectively making all the nodes indistinguishable and eliminating
heterogeneities due to boundary effects.

\begin{figure}[!tbp] \begin{center}
\includegraphics[width=1\textwidth]{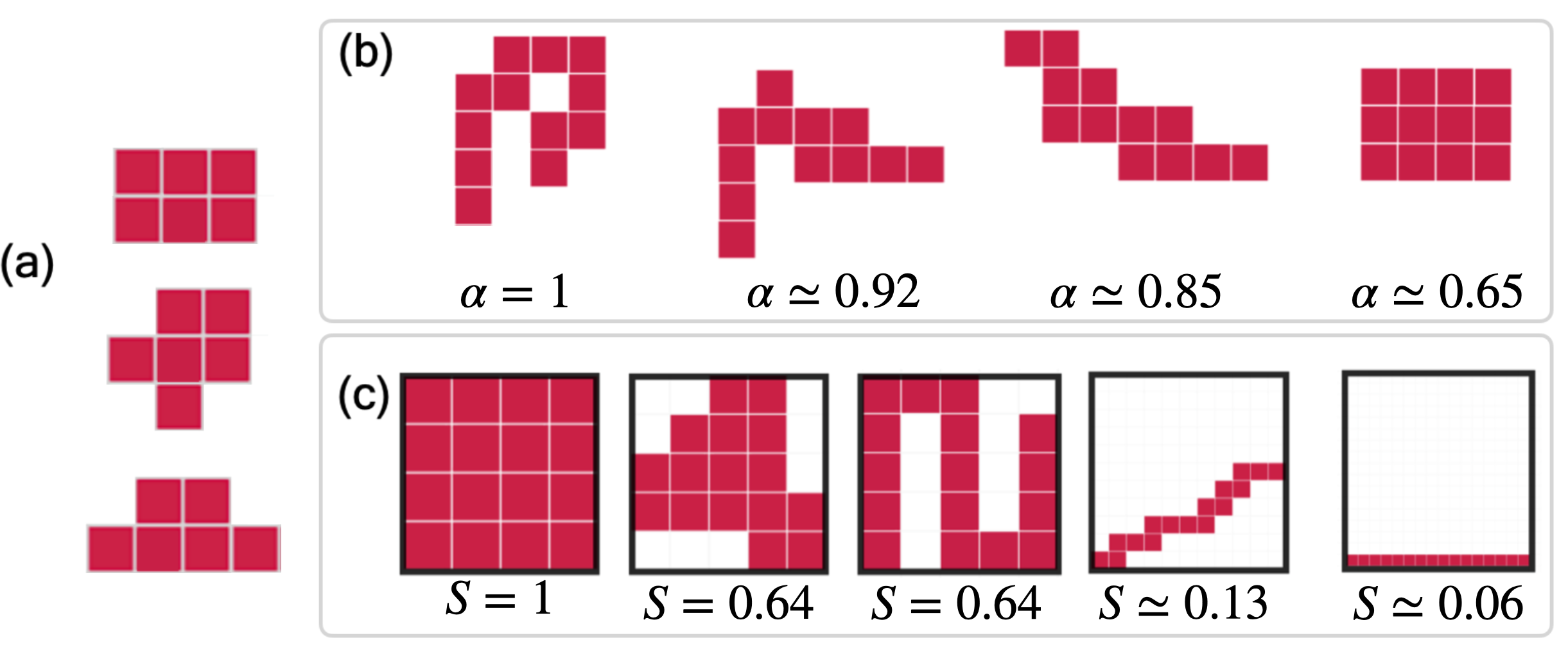} \end{center}
\caption{ (a) Three distinct cluster configurations of the free cluster
with $6$ nodes, where the edges of the cluster are represented by
adjacency between nodes.  (b) Four different cluster configurations of a
free cluster with size $N=12$. The corresponding value of tree-likeness
$\alpha$ is shown right beneath each configuration, and the
configurations are ordered accordingly to their $\alpha$: the leftmost
configuration is the one with the highest value of tree-likeness in the
set. (c) Five different cluster configurations for the same free cluster
with dimension $N=16$, along with their shape factor $S$. The
configuration with the largest value of $S$ in the set is the leftmost
one.} \label{fig_notation} \end{figure}

Each node $i$ of $\mathcal{G}$ is assigned a colour $c_i$, chosen from a
set $C$ of available colours. In order to describe coloured patterns, we
define a free cluster $\mathcal{C}$ as a maximal connected subgraph of
$\mathcal{G}$ characterised by nodes of the same colour $c$.  We call
the size, or dimension, of a free cluster the number $N$ of nodes in
$\mathcal{C}$. The shape of free clusters having the same size $N$ can
vary considerably, depending on the relative position of the nodes
belonging to it. We call cluster configuration, or simply cluster, a
specific spatial arrangement of nodes of a free cluster at a given size
$N$ (see Fig.~\ref{fig_notation}(a)).  In the case of lattices, we state
that two clusters are distinct if their node positions cannot be
obtained one from the other by translation, reflections, or rotation of
node positions (or any combination of these spatial transformations).
Notice that the number of different configurations of a free cluster
with $N$ nodes is obviously finite, but still unknown in the case of
clusters on square lattices for relatively small values of
$N$~\cite{Silva2007, Shirakawa2012, Mason2023}. 

The different cluster configurations associated with a free cluster of
size $N$ can be distinguished by a variety of structural properties. The
most basic one is the number $e$ of edges connecting all the nodes of a
cluster configuration, which obviously depends on the relative positions
of the nodes belonging to the cluster. We define the surface $\sigma$
of a cluster the number of edges between the nodes in the cluster and
the rest of the graph, i.e., all the nodes adjacent to nodes of the
cluster that are of a different colour from that of the cluster.  Among
all the possible cluster configurations with a given size $N$, clusters
with a small surface are indeed the more compact ones, i.e., whose nodes
are more frequently connected to nodes in the cluster, rather than
outside of the cluster.  This preference has an obvious impact on the
shape of a cluster configuration.

Another property of a cluster configuration that is connected to its
structure and geometry is the so-called tree-likeness~\cite{Cook1970,
Potebnia2018}. For the purpose of this work, we define the tree-likeness
$\alpha$ of a cluster configuration as the ratio between $N-1$ and the
number $e$ of its edges. Indeed, the maximum value of $\alpha$ is $1$,
as the cluster is by definition connected, and the minimum number of
edges in a connected graph of $N$ nodes is $N-1$. 

It is easy to show that $\alpha > 0.5$ in the case of square lattice
grids and tori. Since we are in a 2D lattice, which is obviously a
planar graph~\cite{Trudeau1993, Barthelemy2017b}, there is exactly one
edge between any pair of nodes that are adjacent on the lattice (and
only among those nodes that are adjacent in the lattice). The
consideration of finite lattices implies a finite number of edges,
bounded by a value dependent on the lattice's size, and it is related to
the number of edges in the largest possible cluster within that lattice.
For a torus with size $N$, the largest cluster that we can imagine has a
size equal to that of the entire torus, and the number of edges in this
cluster is precisely $2N$. For a grid with the same size, the largest
possible cluster also covers the entire lattice. However, at difference
with the torus, we need to consider the edges that were excluded from
the count due to the torus boundary conditions. So in the end, for the
largest cluster in the grid we have $2N - 2\sqrt{N}$ edges. Since $2N -
2\sqrt{N} < 2N$, we have that the tree-likeness of a generic cluster with
size smaller or equal to $N$ is bounded from below as follows:
\begin{equation} \begin{aligned} \alpha \ge \frac{N-1}{2N}  \\ \end{aligned}
\label{eq:treelikeness} \end{equation}
and for $N \to \infty$ this lower bounds tends to $0.5$. So, in the case
of finite square lattice grids and tori, for a cluster with size $N$,
$\alpha$ takes values in the range $[\frac{1}{2} - \frac{1}{2N}, 1]$.
In Fig.~\ref{fig_notation}(b) we show some cluster configurations and
the associated values of $\alpha$ for $N=12$. Note that values of
tree-likeness close to $1$ indicate stripy or pitted clusters, while more
compact configurations have a value of $\alpha$ close to $0.5$.

Finally, we define the shape factor $S$ of a cluster configuration as
the ratio between the number of nodes in the cluster and the number of
nodes in a pre-defined convex bounding box that contains the cluster
completely~\cite{Harris1964, Polsby1991, Wenwen2022, Montero2020,
Wirth2020}. In this study, we consider as a bounding box the square with
side equal to $D_S = 1+ \max\left\{\max_{i,j}\left\{|x_i-x_j|\right\},
\max_{i,j}\left\{|y_i-y_j|\right\}\right\}$, where $(x_i, y_i)$ are the
lattice coordinates of all the nodes $i$ belonging to the cluster.
In Fig.~\ref{fig_notation}(c) we show different cluster configurations
for $N=16$ with their bounding box and shape factor. At fixed $N$, the
more elongated the cluster, the larger its bounding box and the smaller
the shape factor. For our definition of shape factor, the configuration
with the highest $S$ is indeed the one corresponding to a square with
side $\sqrt{N}$. We loosely call \textit{dense} a cluster with a shape
factor close to $1$, while clusters with a shape factor close to zero
are called \textit{elongated}, for obvious reasons. Similarly, for a
fixed value of $N$, the cluster with the smallest possible shape factor
in a square lattice (i.e., the most \textit{elongated} one), is a line
of nodes all having the same $x$ or $y$ coordinates, which yields
$S=1/N$ (see the rightmost sub-plot of Fig.~\ref{fig_notation}(c).  This
means that the shape factor of such lines effectively tends to $0$ as
$N$ grows.

\section{Colouring processes} Graph colouring is a topic that affects
many different scientific areas, including computer science, operations
research, scheduling, and biology \cite{Kubale2004, Behnamian2016,
Sudev2020}. Indeed, it is only by colouring or labelling nodes according
to their function that we can often recognise the emergence of
interesting patterns in a complex system, where \textit{interesting} is
any structure that hints to the emergence of a spontaneous or
orchestrated organisation~\cite{Papadimitriou2020}.
A colouring over the set of nodes of a graph $\mathcal{G}$ is a function
that assigns to each node one of the available colours or labels in a
set $C$ of discrete elements, with size $|C|$. The assignment can be
performed in many different (and arbitrary) ways and is often a random
process associated with an underlying colour distribution \probP. 

\subsection{The Uniform Random Colouring Process and the Random Growth
Model.} 

We consider here the Uniform Random Colouring (\modurc) process,
as a specific colouring process that assigns to each node of the network
$\mathcal{G}$ one of the available colours by randomly sampling it from
a pre-determined colour distribution \probP. In the \modurc
process, the colour assigned to each node is independent from the colour
of any other node of the graph: given a set $C$ of available colours,
the distribution \probP defines the probability for a node in the
network to be coloured with the colour $c \in C$.  
The \modurc process is indeed the simplest stochastic process
that preserves a given distribution of colours, without introducing
correlations among colours. Since we are interested in assessing the
significance of coloured clusters in a spatial network, and how the
formation of spatial clusters entails the creation of some degree of
correlation in the colours of adjacent nodes,  we propose to use the
\modurc process as a minimalist null-model against which the
importance of correlations and heterogeneity in a spatial network with
coloured nodes can be quantified.

In the following, we study how the structural properties of an
\modurc cluster, including its size, shape, and compactness,
depend on the size of the underlying graph $\mathcal{G}$ and on the
number of available colours. To this aim, it is convenient to consider
an analogous of the \modurc model as a growth process that
obtains  coloured clusters one by one, starting from an initial seed. In
this way, we can obtain many clusters of prescribed size at the same
time. We call this growing process the Random Growth Model
(\modrgm). Growth models are commonly used in many areas of
applied mathematics and physics~\cite{Eden1998, Turner2019, Candia2001,
Jiang1989, Damron2018}, and can provide a framework for understanding
how complex systems evolve and change over time, e.g., in response to
external changes or stimuli. We call a cluster growth process a specific
colouring over $\mathcal{G}$, where the assignment of colours over the
nodes follows a specific  colouring function, i.e. a growth rule, and
the growth starts from a single node, or seed, assigned with a specific
colour $c$, where we mark the seed with the label "$\times$", and all
the other nodes are $blank$, i.e. nodes that have not yet been coloured.
A cluster growth process starts from the first growth step, i.e. $t=0$,
ad it proceeds by applying the growth rules for $t>0$ until the process
stops.
Usually, the rule specifies which adjacent blank nodes are eligible to
be coloured at the next step $t$ by the colouring function, and we call
random growth models such growth processes that assign colours to nodes
in $\mathcal{G}$ following a stochastic colouring function.

\begin{figure}[!htbp] \begin{center}
\includegraphics[width=1\textwidth]{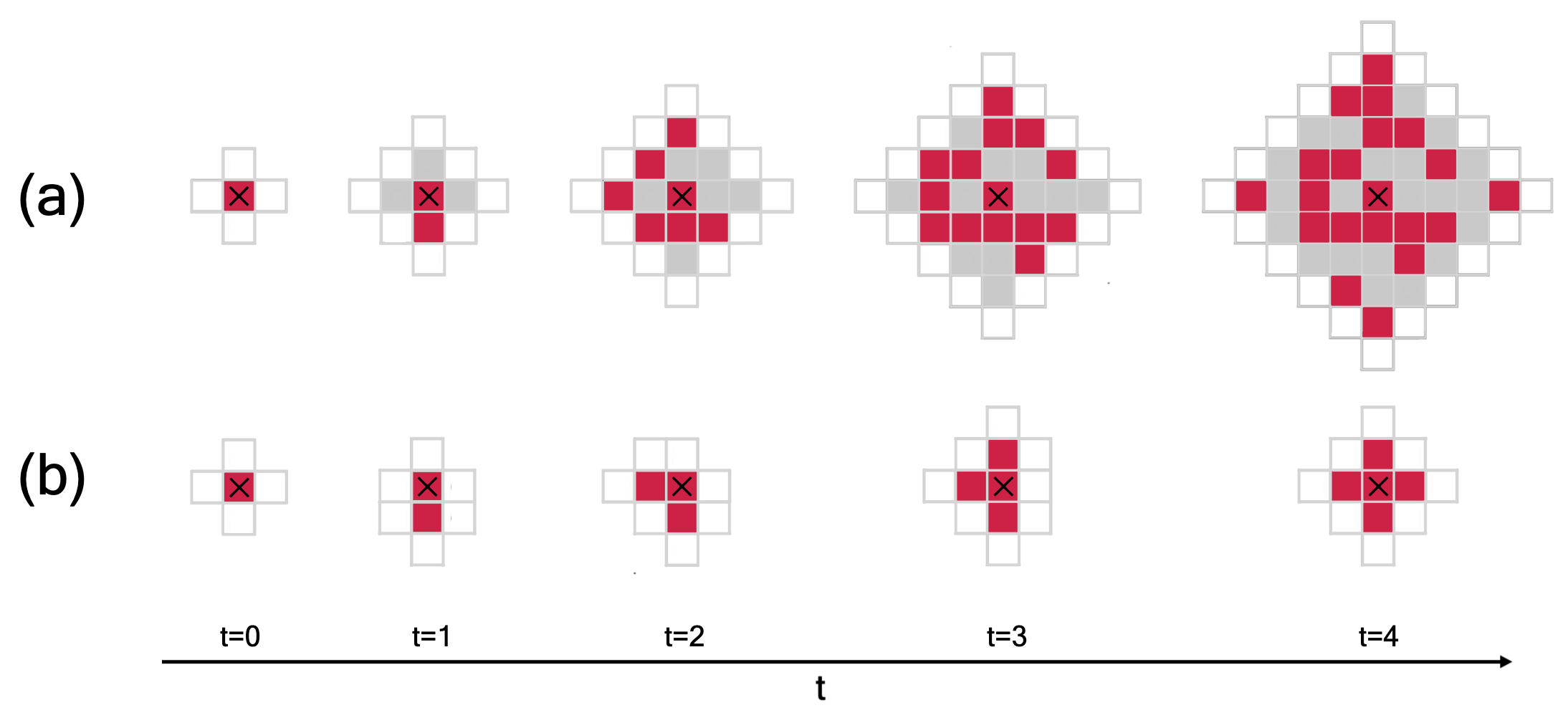}
\end{center}
\caption{ (a) One instance of \modrgm for five growth steps. At $t=0$,
the seed cluster (labelled as  "$\times$"), has only one node and is
surrounded by blank nodes. Here we only show the blank nodes that are
one step away from the coloured cluster. At $t=3$, the seed cluster
grows considerably by acquiring several new nodes.  When all the blank
sites adjacent to the cluster are assigned with a colour different from
the one of the cluster, the cluster can not grow anymore, as it happens
in this case for $t=4$. (b) One instance of \modegm is shown for five
growth steps.  The seed cluster is labelled as "$\times$", while only
blank nodes that are one step away from the coloured subgraph are shown
for each growth step. Note that an \modegm cluster grows by one node at
each step, with probability 1. } \label{fig_rgmegm} \end{figure}

In \modrgm the growth starts from a graph $\mathcal{G}$ where only one
node is assigned with colour $c \in C$, as shown in
Fig.~\ref{fig_rgmegm}(a) in the case where the graph $\mathcal{G}$ is a
squared lattice. This single first node is the seed of our growing
cluster. At each subsequent step $t$, we sample a colour from $C$ for
each of the nodes adjacent to nodes in the cluster, according to the
pre-determined colour distribution \probP.  If at least one of the
adjacent nodes of the current cluster is assigned colour $c$, then the
cluster has grown (since it has acquired at least one node), and the
process can continue.  Otherwise, there is no possibility for the
cluster to grow further, as all the neighbours of the seed are assigned
a colour different from $c$, and the process terminates. Notice that,
since the assignment of colours to nodes is performed in a random and
independent way, still according to the underlying colour distribution
\probP, the ensemble of clusters generated by \modrgm is equivalent to
the ensemble of clusters generated by the \modurc process in a lattice
whose size $N$ goes to infinity. As we will see in the following,
\modrgm is a computationally more convenient way to study the behaviour
of uniform random colouring as their sizes increases, so we will refer
to \modurc and \modrgm interchangeably, as the only difference between
the two models is the actual algorithm used to generate clusters with
them.

\subsection{Correlated clusters and the Eden Growth Model.} As we aim to
characterise coloured patterns, we choose to compare the \modrgm
clusters with the ones produced by a growth model which is very well
known in the literature, as it is largely used to study plant formation
and bacterial growth, and to model a variety of biological systems: the
Eden Growth Model (\modegm). Many variations of this growth process have
been studied for shaping and modelling the natural realms in many
different fields of study~\cite{James2004, Frey2020, Ivanenko1999,
Agyingi2018}. 
Here, we consider the first and most basic formulation of \modegm, which
produces an ensemble of motifs with a limiting shape tending to a circle
when the size $N$ grows to infinity~\cite{Manin2023}. In \modegm, the
growth starts from a blank graph with a coloured seed cluster of one
node labelled with colour $c$, as in the case of \modrgm
(Fig.~\ref{fig_rgmegm}(b)). At each time step $t$, one of the edges
between the seed and the blank nodes is chosen uniformly at random, and
the blank node connected with that chosen edge is assigned with colour
$c$ with probability $p=1$.  Notice that this rule forces a cluster to
continue growing indefinitely, as each step of the model adds a new node
to the existing cluster. This is fundamentally different from the case
of \modrgm, as in that case the growth might die at any step.

Despite its elegant simplicity, the first formulation of the Eden Growth
Model is still extensively used in many scientific fields, including
urban growth and biology, to model the emergence of circle-like spatial
arrangements \cite{Teknomo2005, Waclaw2015, Santalla2018}. In this
model, the probability for a node to acquire a given colour does depend
on the colour of its neighbours, and on the diffusion rule. Hence, the
clusters generated by \modegm exhibit quite strong spatial
correlations.  We will mainly use the \modegm process to test the
robustness and descriptiveness of our measures and also to make
comparisons with the clusters obtained with the \modrgm growth,
which instead are, by definition, uncorrelated.

\section{Results}
\subsection{Cluster size on grid and torus.} 
\begin{figure}[!tbp]
\begin{center} \includegraphics[width=1\textwidth]{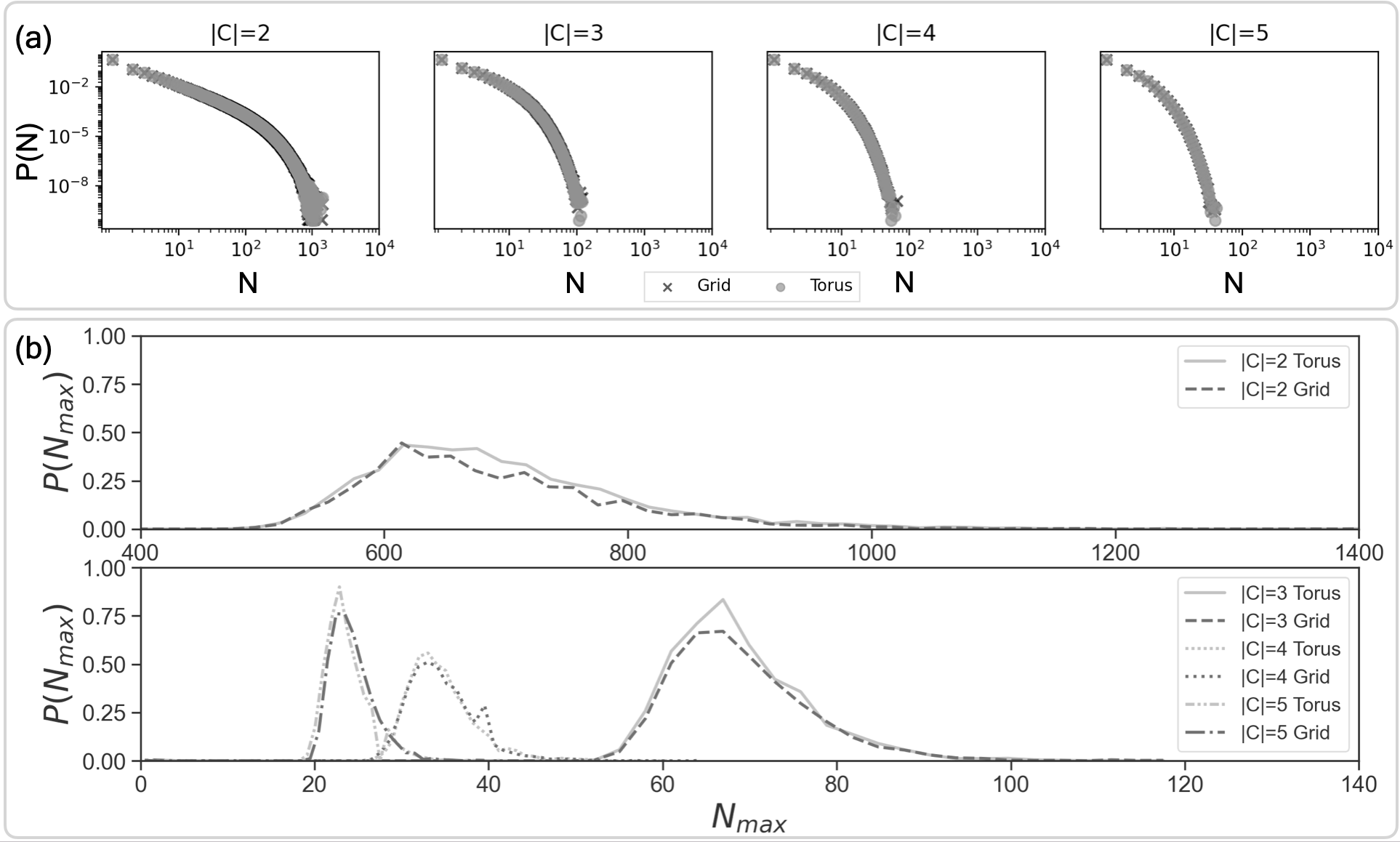}
\end{center}
\caption{ (a) Probability distribution $P(N)$ of the cluster size $N$
for a \modurc process run over a grid (dark grey) and a torus
(light grey), for different values of $|C|$. (b) Probability
distribution $P(N_{max})$ of the size $N_{max}$ of the largest cluster
in the \modurc ensemble on a grid (dark grey) and a torus (light grey)
with the same dimension.  The case $|C|=2$ is shown in the top panel,
while in the panel below the cases $|C|=3$, $|C|=4$, and $|C|=5$ are
depicted. Notice that the number of colours has an impact on both the
distribution of sizes and, more importantly, on the size of the largest
cluster. All the distributions are computed over $5\times 10^4$
realisation of he corresponding model over tori or grids with  $4 \times
10^6$ nodes.} \label{fig_gridtorus} \end{figure}

Here we show the properties
of coloured clusters obtained by the \modurc model on finite
grids and tori. We start from a uniform distribution of colours, meaning
that the probability of assigning colour $c$ to each node is equal to
$p=1/|C|$. We ran a large number ($5\times 10^4$) of simulations of the
\modurc process on each type of graph, collecting the size of all
the clusters across all the realisations. For our study, we choose
graphs with size $4 \times 10^6$.
As we can see from Fig.~\ref{fig_gridtorus}(a), for $|C|=2$ we have a
higher chance of observing clusters with a larger size than in the case
of a different number of available colours. For $|C|=3$, the largest
size observable is around $10^2$, and for $|C|=4$ and $|C|=5$ the
largest size decreases even more. These results are in agreement with
percolation theory on square lattices~\cite{Newman2000, Mertens2022}.
So, except for the singular case of $|C|=2$, we can conclude that the
probability of observing extensive clusters on a grid or a torus is
indeed an exponentially decreasing function of the cluster size $N$. 

\subsection{The largest cluster in the \modurc process.} 

The size of the largest cluster obtainable on a lattice of a given size
remains a subject of limited knowledge. Indeed, the size of a maximal
cluster depends on a multitude of factors that have yet to be fully
comprehended~\cite{Zierenberg2017}. We now discuss the results of
$5\times 10^4$ numerical simulations to obtain the distribution of the
size of the largest \modurc cluster on a lattice. We obtained the
largest cluster size $N_{max}$ for clusters on both grid and torus with
side $D=2000$ (i.e., with $4 \times 10^6$ nodes), and we look at the
distribution of $N_{max}$ for different values of $|C|$.  The position
and the value of the peak of the distributions strictly depend on $|C|$
(Fig.~\ref{fig_gridtorus}(b)).  The higher the value of $|C|$, the more
the peak appears to shift to the left, meaning that the probability of
obtaining large clusters in a \modurc with a large number of
colours is smaller. The size of these clusters depends on the number of
colours available, and it becomes quite irrelevant when compared to the
overall size of the torus when $|C|$ increases.  Despite the largest
\modurc cluster being normally smaller than the overall graph,
there remains a substantial probability of encountering clusters with
relatively significant size, especially for $|C|=2$ (see
Fig.~\ref{fig_gridtorus}(b)). 
This is indeed the most interesting scenario in the case of a
\modurc process, as collecting large clusters in this case is
somehow easier.  Therefore, in the following we will only focus on
clusters generated with a \modurc colouring in which $|C|=2$ and
the probability of assigning one of the two available colours (namely
$c$ and $c*$) is the same and equal to $p=\frac{1}{2}$.

\subsection{Tree-likeness and Shapefactor}

\begin{figure}[!tbp] \centering
\includegraphics[width=1\textwidth]{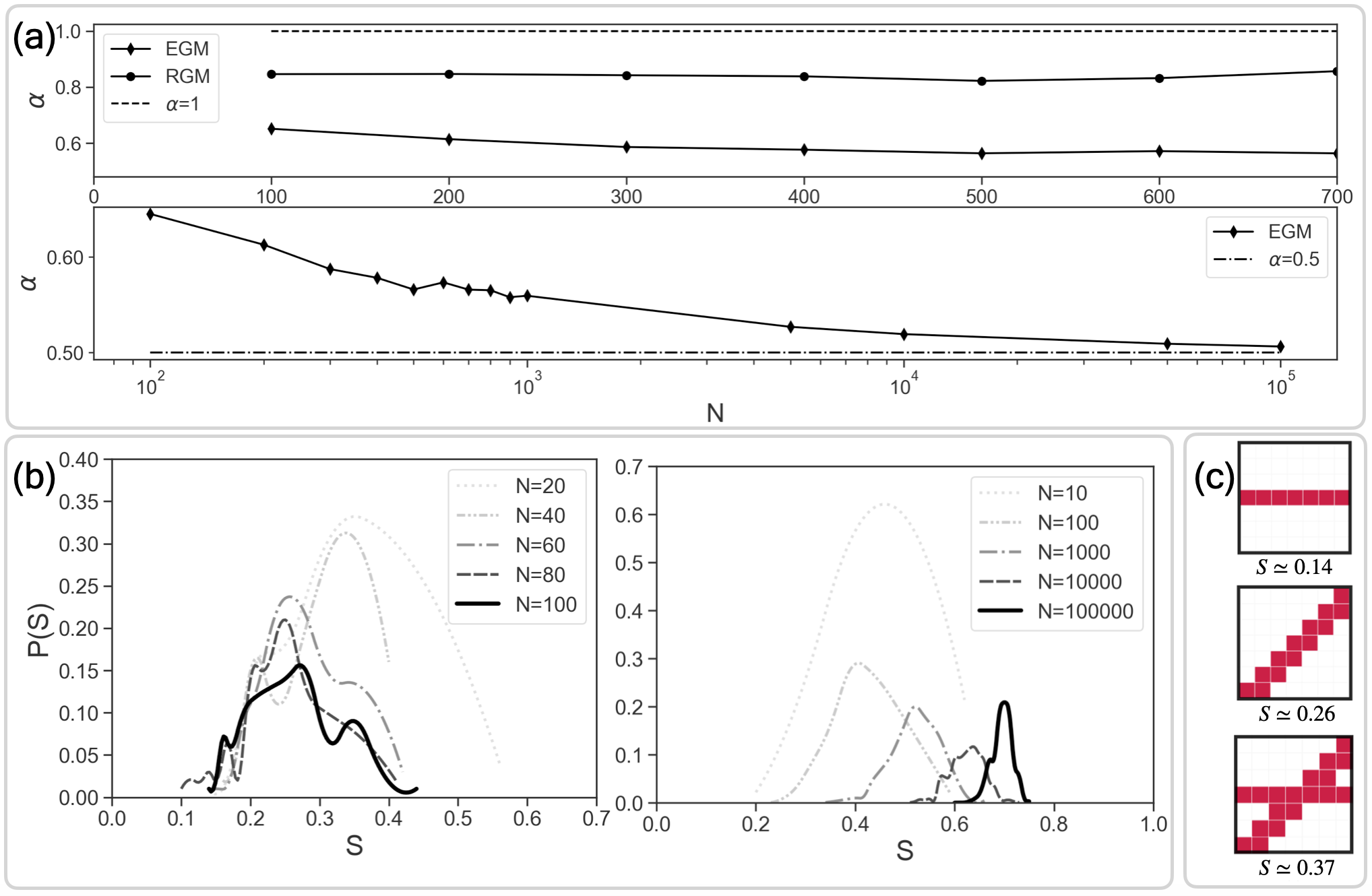} \caption{ (a)
Values of tree-likeness $\alpha$ for \modrgm and \modegm clusters for $N$
in the range $N=[100,700]$ (top) and for \modegm clusters (bottom) for
$N\in[10^2, 10^5]$. The two panels also show the reference values
$\alpha =1$ and $\alpha=0.5$. (b) Shape factor distribution $P(S)$ for
\modrgm (left) and \modegm (right) clusters.  (c) The three typical and
more probable local arrangements of \modrgm clusters, where the side of
the domains in this case is $L=7$. The upper and the central domains
represent two configurations in which the nodes are distributed as
stripes, one is the stripe with the lowest value of shape factor (top),
and one is the stripe along the diagonal of the square domain (middle).
The last domain (bottom) represents the union between the two previous
configurations, with a higher value for the associated shape factor. All
the data points are obtained over $10^5$ cluster realisations for each
value of $N$. } \label{fig_treelshape} \end{figure}

In Fig.~\ref{fig_treelshape}(a), we show the tree-likeness $\alpha$ for
\modrgm and \modegm clusters in the size range
$N=[100,700]$ (top) and for the \modegm clusters (bottom) with
$N$ up to $10^5$. 
As shown in the top panel of Fig.~\ref{fig_treelshape}(a),
\modrgm clusters present a tree-likeness close to $0.8$, which is
indicative of an essentially stripy shape. On the other hand, the
\modegm clusters exhibit smaller values of $\alpha$, reaching the
limit value of $\alpha=0.5$ for large $N$ (see
Fig.~\ref{fig_treelshape}(a), bottom) as expected for the compact,
filled, circle-shaped structures generated by the Eden Growth model. 

In Fig.~\ref{fig_treelshape}(b), we report the distribution of cluster
shape factor over $10^5$ realisations of \modrgm and
\modegm clusters for different values of $N$. The peak of the
distribution shifts to the left for the \modrgm clusters when $N$
increases.  This means that \modrgm generates elongated clusters
rather than dense ones as their size grows. On the contrary, in the case
of \modegm, the peak shifts in the opposite direction, meaning
that the shape factor is increasing with size, and reaching the value of
shape factor for the limiting shape of \modegm clusters, which is
the one of a circle. In fact, for a circle-like shape in a bounding box with side $2R$, we have that the shape factor is equal to:
\begin{equation*}
S=\frac{\pi R^2}{(2R)^2} = \frac{\pi}{4} \simeq
0.78
\end{equation*}
which is around the value of the shape factor we measured for the peak
of the \modegm clusters at $N=10^5$. So, the shape factor is
correctly telling us that \modegm clusters shape is tending to
the one of a circle when $N$ increases.

The difference in the range of cluster sizes between the two models, as
made evident in Fig.~\ref{fig_treelshape}(b), is due to the different
rates at which clusters with a fixed dimension $N$ are generated: in the
case of \modrgm clusters, because of the lower probability in observing
large clusters, generating a statistically significant number of
clusters in the dimension range $N=[10^3,10^5]$ is comparatively harder
- so the need to find another approach for describing the limiting shape
  of these objects.

We can derive a simple mean-field approximation for the limiting shape
factor of \modrgm clusters, based only on the analysis of its
spatial distribution at a smaller scale. 
Let us divide an \modrgm cluster into a set of non-overlapping
square domains, all with the same side $L$, where $L \ll D_S$  and $D_S$
is the side of the bounding box of the cluster.  Thanks to the results
in Fig.~\ref{fig_treelshape}(b), we know that a \modrgm cluster
is not dense, but at the same time, the peak value of the shape factor
for \modrgm with $N=100$ is between $0.2$ and $0.3$, so higher
than $1/N$, which means that they are not in the most elongated
configuration.  
By looking at the shape of \modrgm clusters, we make the
reasonable assumption that each of the sections of the cluster after the
subdivision will be similar to one of the three predominant shapes shown
in Fig.~\ref{fig_treelshape}(c) for the case where $L=7$. The domains in
Fig.~\ref{fig_treelshape}(c) top and middle represent two stripy
configurations of nodes: the first is the one with the smallest shape
factor for that $L$, while the second is the configuration of nodes over
the diagonal of the square with side $L$. The domain in
Fig.~\ref{fig_treelshape}(c) bottom represents the crossing of two of
the two shapes described above. If we make the mean-field assumption
that the abundance of these three domains is the same in the subdivided
cluster, we have that the overall shape factor has to be:
\begin{equation} S = \frac{\frac{7}{49} + \frac{13}{49} + \frac{18}{49}}{3} \simeq \frac{0.14+0.26+0.37}{3} \simeq 0.26 \\ \label{eq:shape
factor_MF} \end{equation}
Which is comparable to the value for the peak shape factor  we measured
for the \modrgm clusters.

\subsection{Mean Exit Time from a cluster}

The size, tree-likeness and shape factor of a cluster provide some useful
hints about its geometry, but it is difficult to condense that
information in a single structural indicator. Here we propose to use the
expected time needed for a random walker to leave the cluster, also
known as hitting time~\cite{Bernt2003} or exit time, as a comprehensive
descriptor of the geometry of a cluster. Let us consider a time-discrete
uniform random walk on $\mathcal{G}$~\cite{Masuda2017}, such that the
one-step probability for a walker sitting at node $i$ with degree $k_i$
to jump to one of the neighbours $j$ of $i$ is equal to $\pi_{ij} =
1/k_i$.  Notice that with this notation the transition matrix
$\Pi=\left\{\pi_{k\ell}\right\}$ associated to the walk is
row-stochastic. The hitting time $\tau_{i,j}$ from node $i$ to node $j$
is the expected number of steps needed for a random walk starting at $i$
to reach node $j$ for the first time. The recurrent forward master
equation for $\tau_{i,j}$ can be written as: 
\begin{equation}
\tau_{i,j} = 1 + \sum_{k \in \mathcal{N}} \pi_{ik}\tau_{k,j}
\end{equation}
The hitting time is a measure of how difficult it is to reach node $j$
by means of an unbiased diffusion process started at $i$: the higher the
value of $\tau_{i,j}$, the more remote $j$ is from $i$. We define the
frontier $\fC$ of the cluster $\mathcal{C}$ as the set of nodes of
$\mathcal{G}$ which have at least one edge connecting them to a node in
$\mathcal{C}$ but whose colour is different from the colour of the
cluster $\mathcal{C}$. We define the Mean Exit time $\tau_{\mathcal{C}}$
from the cluster $\mathcal{C}$ with $N$ nodes as the average time needed
to a walker to reach the frontier of $\mathcal{C}$ when starting from
one of the nodes of $\mathcal{C}$. In other words, $\tau_{\mathcal{C}}$
is the average of $\tau_{i,j}$ for all the nodes $(i,j)$ such that $i\in
\mathcal{C}$ and $j\in \fC$, in formula:
\begin{equation} 
\tau_{\mathcal{C}} = \frac{1}{N|f_{\mathcal{C}}|}
\sum_{i\in\mathcal{C}}\sum_{j\in\fC}\tau_{i,j}
\label{eq:eq_tau_c} 
\end{equation}  
Notice that, in general, the frontier of the cluster $\fC$ contains more
than just one node. According to our definition of exit time from a
cluster, we are interested in the average time needed for a walker to
reach for the first time any node not belonging to the cluster,
independently of the actual node reached by the walker. For this reason,
in the following we will denote by $\tau_{i,\fC}$ the average exit time
for walkers starting at node $i$ in the cluster $\mathcal{C}$, i.e., the
quantity $\tau_{i, f_{\mathcal{C}}}=\frac{1}{|\fC|}\sum_{j\in \fC}
\tau_{i,j}$. The computation of the mean exit time from a subgraph of
$\mathcal{G}$ can be easily formulated as a linear system that depends
only on the walk transition matrix $\Pi$ (and, consequently, on the
structure of the graph), as shown in Ref.~\cite{Bassolas2021bis}.
However, for very large graphs (anything with more than about $10^7$
nodes) the solution to that system of equations can become
computationally unfeasible. In those cases, a suitable approximation of
$\tau_{\mathcal{C}}$ can be obtained by simulating a large
number of walks starting from each of the nodes of the cluster.

\begin{figure}[!htbp]
\begin{center}
\begin{subfigure}[][][t]{2.8in}
\includegraphics[width=4in]{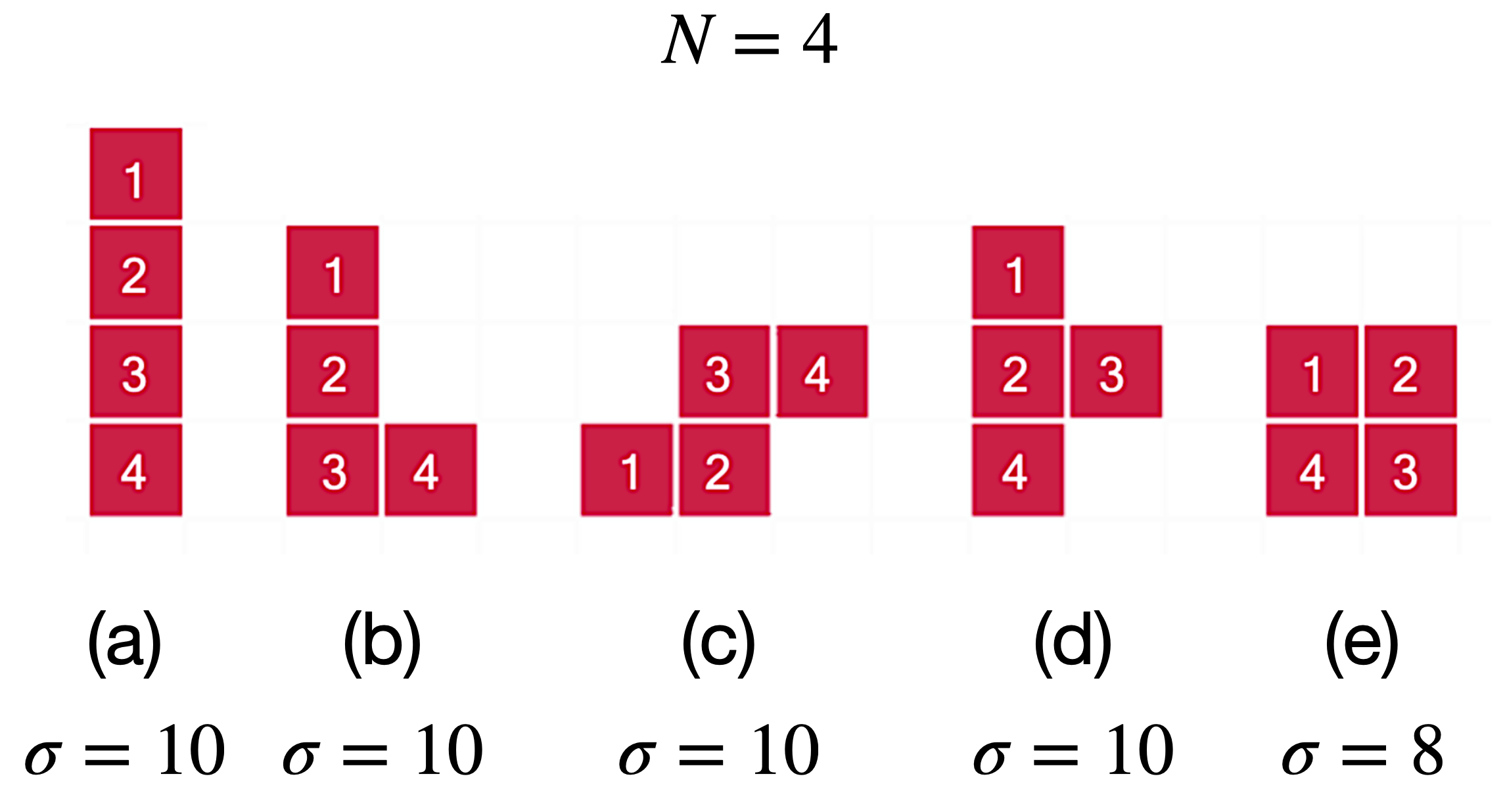}
\end{subfigure}
\hfill
\begin{subfigure}[][][t]{3in}
	\begin{tabular}{c|c|c|c|c}
	 configuration & $\tau_{\mathcal{C}}$ & $\sigma$ & $\alpha$ & $S$ \\  \hline \hline
	(a) & $1.\overline{63}$ & $10$ & $1$ & $0.25$ \\
   	(b) & $1.\overline{63}$ & $10$ & $1$ & $0.45$ \\
        (c) & $1.\overline{63}$ & $10$ & $1$ & $0.45$ \\
	(d) & $1.69$ & $10$ & 1 & $0.45$ \\
	(e) & $2$ & $8$ & $0.75$ & $1$\\
	 \hline
	 \hline
	\end{tabular}
\end{subfigure}
\end{center}
\caption{ (a)-(e) All the five cluster configurations of a free cluster
with size $N=4$. For each of the configurations, we also show the value
of the surface $\sigma$. The table shows the mean exit time, surface,
tree-likeness, and shape factor for each of the five cluster
configurations.} 
\label{fig_N4} 
\end{figure}

The mean exit time for different cluster configurations with the same
size does indeed depend on several geometric properties of the cluster
configuration, including its shape, tree-likeness, depth (that is, the
average distance from each node to the frontier of the cluster), and so
on. As a simple example, we consider the case of clusters with $N=4$
nodes, and in Fig.~\ref{fig_N4}, we show all the possible cluster
configurations for the size $N=4$, along with their surface $\sigma$. 

The exit time from each of the nodes in the cluster configuration in
fig.~\ref{fig_N4}(a) is obtained by solving the following system of
equations:
\begin{equation*} \begin{array}{rcl}
\tau_{1,\fC}&=&1+\pi_{12}\tau_{2,\fC}  \\
\tau_{2,\fC}&=&1+\pi_{21}\tau_{1,\fC}+\pi_{23}\tau_{3, \fC}\\
\tau_{3,\fC}&=&1+\pi_{32}\tau_{2,\fC}+\pi_{34}\tau_{4, \fC}\\
\tau_{4,\fC}&=&1+\pi_{43}\tau_{3,\fC}  \\
\label{eq:first_4block_firststep} \end{array} \end{equation*}

Where $\pi_{ij}$ represents the probability that a walker at node $i$
jumps to node $j$ in one step. In our case, all the one-step
probabilities are the same and equal to $1/4$, i.e., as each node in the
lattice is connected with exactly four neighbours. 
If we solve this system of equations, we obtain $\tau_{1, \fC} =
\tau_{4, \fC} = 16/11$ and $\tau_{2, \fC}= \tau_{3, \fC} = 20/11$. This
reflects the fact that the cluster has a rotational symmetry, which makes
the two pairs of nodes $(1,4)$ and $(2,3)$ geometrically equivalent.
Hence, for this configuration, we obtain the mean exit time from the
cluster:
\begin{equation*} \tau_{\mathcal{C}}^{(a)}= \frac{ 2 \times
\frac{16}{11}+2 \times \frac{20}{11}}{4} =\frac{18}{11} = 1.\overline{63} 
\label{eq:first_4block_meanexit} \end{equation*}
It is easy to show that the configurations shown in fig.~\ref{fig_N4}(b)
and fig.\ref{fig_N4}(c) lead to the same system of equations, so the
exit time from those clusters is the same as that of the cluster in
fig.~\ref{fig_N4}(a), i.e.,
$\tau_{\mathcal{C}}^{(a)}=\tau_{\mathcal{C}}^{(b)}=\tau_{\mathcal{C}}^{(c)}$.

Conversely, the cluster in Fig.~\ref{fig_N4}(d) has a different set of
symmetries from the previous ones, which are associated with the following
system of equations:
\begin{equation*}
	\begin{array}{rcl}
		\tau_{1, \fC}& = &1+\frac{1}{4}\tau_{2, \fC}  \\
		\tau_{2, \fC}& = &1+\frac{1}{4}\left(\tau_{1, \fC}+\tau_{3,
			\fC}+\tau_{4, \fC}\right) \\
		\tau_{3, \fC}& = &1+\frac{1}{4}\tau_{2, \fC}  \\ 
		\tau_{4, \fC}& = &1+\frac{1}{4}\tau_{2, \fC}  \\
		\label{eq:second_4block_firststep} 
	\end{array}
\end{equation*}

From that system of equations we can conclude that $\tau_{1, \fC} =
\tau_{3, \fC} = \tau_{4,\fC}$, so we can simplify our calculations
further:
\begin{equation*} \begin{aligned} \tau_{1, \fC}=1+\frac{1}{4}\tau_{2, f}  \\
\tau_{2, \fC}=1+\frac{3}{4}\tau_{1, \fC} \label{eq:second_4block_secondstep}
\end{aligned} \end{equation*}
to obtain the solution $\tau_{1, \fC} = \tau_{3,\fC} = \tau_{4, \fC} =
20/13$ and $\tau_{2, \fC}= 28/13$, which yields the average exit time
from the cluster:
\begin{equation*} \tau_{\mathcal{C}}^{(d)}= \frac{\frac{28}{13}+3 \times
\frac{20}{13}}{4} = \frac{22}{13} \simeq 1.69\label{eq:second_4block_meanexit}
\end{equation*}

Finally, the cluster in Fig.~\ref{fig_N4}(e) is a square. Due to the
symmetric structure of this cluster configuration, we have that
$\tau_{1,\fC} = \tau_{2,\fC} = \tau_{3,\fC} = \tau_{4,\fC}$. By solving
the equation:
\begin{equation*} \begin{aligned} \tau_{1, \fC}=1+\frac{1}{2}\tau_{1,
\fC}\Longrightarrow \tau_{1,\fC} = 2 \end{aligned} \end{equation*}
we obtain ${\tau_{\mathcal{C}}}^{(e)}= 2$. It is interesting to note
that the square in Fig.\ref{fig_N4}(e) is the cluster configuration with
the largest value of exit time. Incidentally, this is also the
configuration with the smallest surface ($\sigma=8$), and the smallest
tree-likeness $\alpha=0.75$. A summary of the geometric properties and
exit times for all the cluster configurations with $N=4$ is shown in the
Table of Fig.~\ref{fig_N4}. In general, the exit time varies even for
configurations having the same size and surface. For instance, the
clusters (a)-(d) all have surface $\sigma=10$ and tree-likeness
$\alpha=1$, but configuration (d) still has a slightly larger value of
exit time. This is intuitively due to the fact that in (a)-(c) all the
nodes have at least two edges to the frontier, while in (d) node $2$ has
only one link to the frontier. As a consequence, a walker starting from
node $2$ in (d) will take comparatively more time to exit from the
cluster, and this results in a slightly larger value of
$\tau_{\mathcal{C}}$. Moreover, the cluster with the largest exit time
is the one with minimal surface, in agreement with the intuition that
the exit time is a proxy for the overall difficulty in leaving a
cluster. In general, it looks like there is no single geometric property
that can predict the exit time alone. Rather, the exit time somehow
summarises a variety of geometric properties of a cluster.

\subsection{Exit time from rectangular clusters.} 
\begin{figure}[!htbp] \begin{center}
\includegraphics[width=1\textwidth]{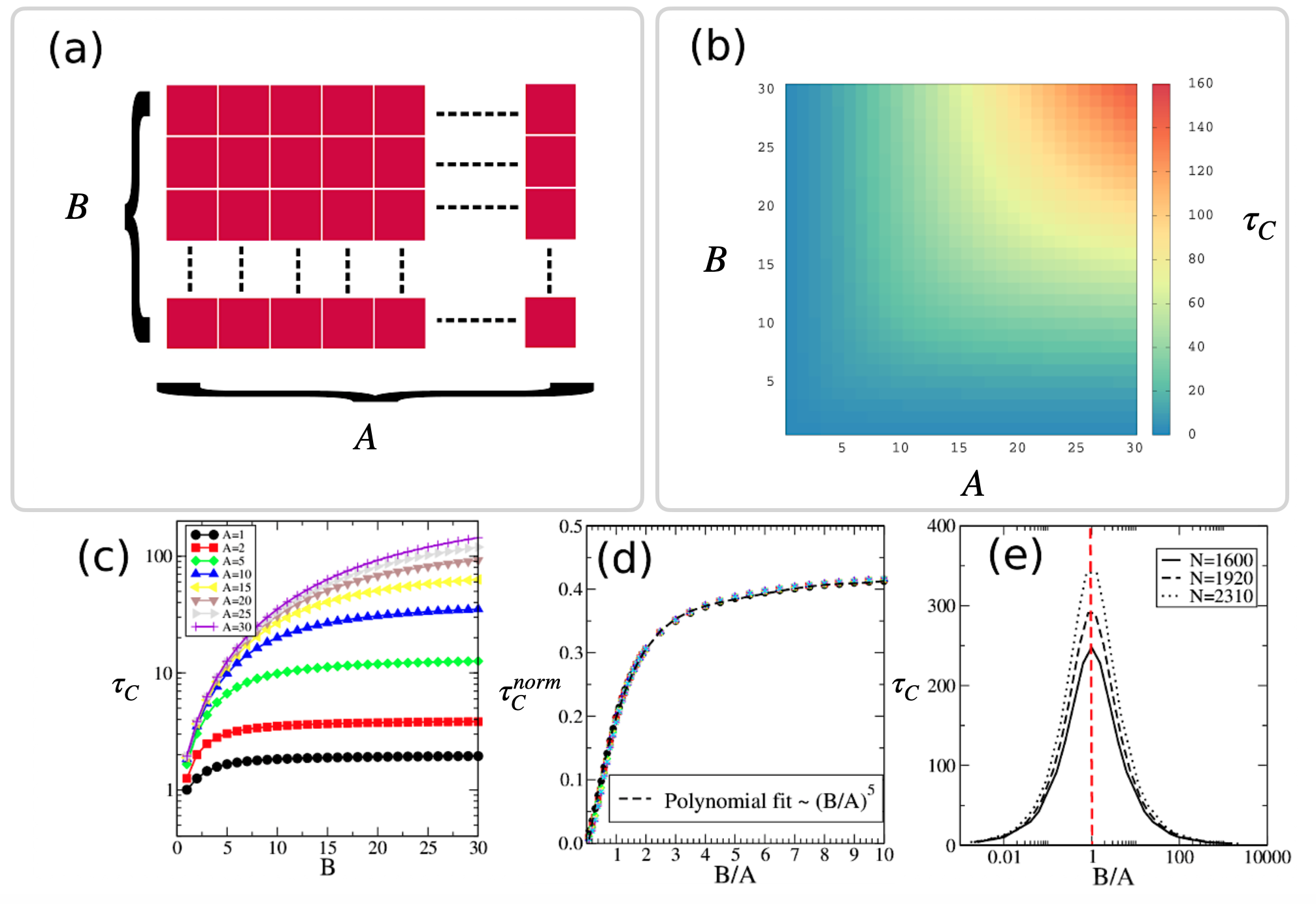} 
\caption{(a) Rectangular clusters with $N=A\times B$ nodes provide
insight on the relation between shape and mean exit time
$\tau_{\mathcal{C}}$. (b)-(c) If we keep the side $A$ fixed and vary the
other side $B$, we observe that the mean exit time is a monotonous
function of $B$.  (d) The $\tau_{\mathcal{C}}(B|A)$ curves for different
values of $A$ can be collapsed on a universal function of $B/A$, where
the normalised exit time $\tau_{\mathcal{C}}^{\rm norm}$ follows very
closely a fifth-order polynomial in $B/A$. (e) If we fix the number of
nodes $N$ in the cluster, and consider all the rectangular clusters
having $N$ nodes, then the maximum of $\tau_{\mathcal{C}}$ is obtained
when the two sides $A$ and $B$ are identical, i.e., for the square
cluster. Moreover, the cluster with the minimal value of $\tau$ is
indeed the line of $N$ nodes, as $\tau_{\mathcal{C}}$ admits only one
maximum for $B/A>0$. } \label{fig:rectangle}
\end{center} \end{figure}

The insight
provided by clusters with $N=4$ suggests that the surface, tree-likeness
and shape of a cluster configuration indeed have a central role in
determining the value of the exit time from that cluster. In an attempt
to collect more evidence about the salient properties of exit times from
compact clusters, here we propose a numerical characterisation of the
exit time from rectangular clusters with $N=A\times B$ nodes (see
Fig.~\ref{fig:rectangle}(a)), and we show that the rectangular
configurations with minimal and maximal exit time for a fixed size $N$
are, respectively, the $1\times N$ rectangle and the $\sqrt{N}\times
\sqrt{N}$ square.  

We start by noting that the exit time from a rectangular cluster is an
increasing function of $B$ for fixed $A$, meaning that in general the
addition of a new row of $A$ nodes to a $(B-1)\times A$ rectangle makes
it more difficult for a random walker to leave the cluster. This is easy
to explain: by adding a new row to the cluster, some of the nodes which
previously were directly connected to the frontier will only have other
cluster nodes as neighbours, which causes an increase of their exit
time.  In Fig.~\ref{fig:rectangle}(b)-(c) we show the exit time
$\tau_{\mathcal{C}}(B|A)$ of a rectangle of sides $A$ and $B$, when one
of the two sides (A) is kept fixed and the other one (B) increases. It
is interesting to note in Fig.~\ref{fig:rectangle}(c) that
$\tau_{\mathcal{C}}(B|A)$ is an increasing non-linear function of $B$
which, for very large values of $B$, saturates to a specific value that
depends only on the other side $A$.  For instance, for $A=1$ we have
that $\tau_{\mathcal{C}}(B|A)$ tends to $2$ when $B$ increases.

Despite we were not able to find a precise analytical expression to
obtain $\tau_{\mathcal{C}}(B|A)$ as a function of $A$ and $B$ only,
there is no doubt that the exit time of a rectangular cluster in a
square lattice only depends on $A$ and $B$, as also suggested by the
existence of a similar qualitative behaviour in
Fig.~\ref{fig:rectangle}(c) for all the values of $A$. Indeed, we were
able to find numerically a normalisation which allows to collapse all
the curves $\tau_{\mathcal{C}}(B|A)$ into a single universal curve,
which is shown in Fig.~\ref{fig:rectangle}(d). The normalisation is
obtained by rescaling the horizontal axis by $A$ and operating the
substitution $\tau_{\mathcal{C}}(B|A) \longrightarrow
\tau_{\mathcal{C}}^{\rm norm}(B/A) =
\frac{\tau_{\mathcal{C}}(B/A)c(A)}{A^2}$, where $c(A)$ is a $3^{\rm
rd}$-degree polynomial in $A$. Notice that, incidentally, $B/A$ indeed
corresponds to the shape factor $S$ of the rectangle for $B<A$, and to
$1/S$ for $B>A$. Hence, this normalisation essentially relates the exit
time from rectangular clusters to their shape factor, in agreement with
the fact that the two dimensions of a rectangle fully determine its
geometry.

The best fit of $\tau_{\mathcal{C}}^{\rm norm}(B/A)$ for values of $A$
and $B$ such that $N\in [10, 10^5]$ (symbols in
Fig.~\ref{fig:rectangle}(d)) is a $5^{\rm th}$-degree polynomial in
$B/A$. Notice that the collapse is perfect over more than four orders of
magnitude. All the curves increase monotonically and indeed saturate for
large values of $B/A$, i.e., for clusters where one of the two sides is
much larger than the other one. 

Finally, in Fig.~\ref{fig:rectangle}(e) we explore how
$\tau_{\mathcal{C}}$ varies when the shape of the cluster is changed and
$N$ is kept fixed. In particular, we considered all the feasible
$A\times B$ rectangles having the same total number of nodes $N$, and we
solved Eq.~\ref{eq:eq_tau_c} to find the corresponding exit time from
the cluster. We report in the figure three values of $N$, namely $1600,
1920, 2310$, of which only $N=1600$ allows for a square cluster.  Notice
that the three curves are concave downward (convex).  Interestingly, the
maximum of $\tau_{C}$ for $N=1600$ is obtained for $B/A=1$, which
corresponds to the square of side $A=B=40$, while the minimum value of
$\tau_{C}$ is obtained for $B/A=1/N$ (and also for $B/A=1600$,
obviously), which corresponds to the $1\times N$ rectangle. A
qualitatively similar behaviour is observed  for the other two curves,
which exhibit two consecutive maximal points just below and just above
$B/A=1$, as those two values of $N$ do not admit a square configuration.   

\subsection{Clusters with minimal and maximal exit times.}

\begin{figure}[!tbp] \centering
\includegraphics[width=1\textwidth]{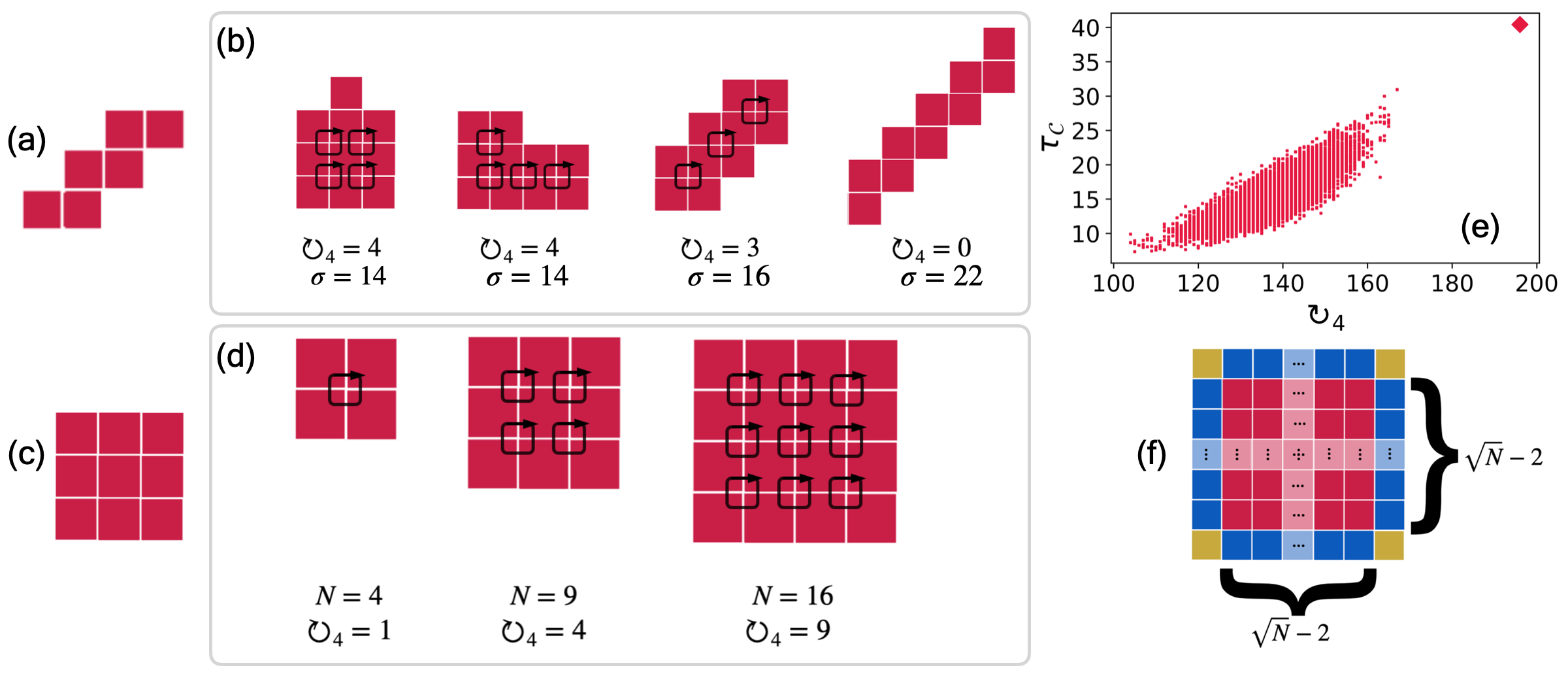} \caption{ (a)
A $line$ cluster configuration for a free cluster with $N=6$. (b) Four
clusters configurations for the same free cluster with $N=10$ are shown,
along with their surfaces and number of simple 4-cycles. Each simple
4-cycle is depicted in the figure as a circle arrow. Configurations with
the lowest value of the surface in the set are on the leftmost part of
the panel. (c)  A $square$ cluster configuration for a free cluster with
$N=9$. (d) Three square configurations, along with their number of
simple 4-cycles and sizes. (e) At fixed $N=225$, $\tau_{\mathcal{C}}$
versus  $\circlearrowright_{4}$ for $10^4$ \modegm realisations (dotted
dataset) and for the square (diamond point) are shown. The square
cluster is associated to the largest exit time of all. (f) A square with
size $N$. The 4 yellow nodes contribute to the number of edges by 2; the
$\sqrt{N}-2$ nodes on the four blue borders contribute with 3 edges; the
$\sqrt{N}-2 \times \sqrt{N}-2$ nodes in the red bulk contribute with 4
edges. For the sake of simplicity, the figure shows only some nodes in a
darker tone, while the dotted light-coloured nodes represent all the
nodes in between. } \label{fig_squareslines} \end{figure}

The analysis of exit times from rectangular clusters has given us
important hints about the dependence of the exit time on the shape
factor of a cluster: the more elongated the cluster (smaller shape
factor, corresponding to "linear" clusters) the smaller the exit time.
Conversely, clusters having a higher shape factor tend to have a higher
exit time, when the size of the cluster is kept constant, with a maximum
reached for square configurations.  Although in general clusters are not
rectangular (and not even convex), an interesting aspect of rectangular
clusters is that their shape factor is intimately connected to the size
of their surface $\sigma$. Indeed, the surface of an $A\times B$
rectangular cluster is $\sigma=2AB$, i.e., equal to the perimeter of the
rectangle.  Hence, of all the rectangular clusters with $N$ nodes, those
having minimal and maximal surface are, respectively, the square of side
$\sqrt{N}$, which has $\sigma=4\sqrt{N}$  and the $1\times N$ rectangle,
whose surface is $\sigma=2N+2$. In the following, we call the $1\times
N$ cluster a "line" (see Fig.~\ref{fig_squareslines}(a)). Notice that in
the specific case of rectangles, surface and exit time are negatively
correlated. Moreover, a rectangle with a larger number of nodes tends to
have a larger exit time (as shown in Fig.~\ref{fig:rectangle}(c)-(d)).
Following these intuitions, we can obtain a simple mean-field
approximation for the exit time from a generic cluster of size $N$  and
surface $\sigma$ as follows: 
\begin{equation} \begin{aligned} \tau_{MF}=\frac{Nk}{\sigma}.
\end{aligned} \label{eq:MF_exittime} \end{equation}
The numerator accounts for the total number of ways in which a walker
can get out of any node of the cluster in one step. This is equal to
twice the total number of edges incident on nodes of the cluster, and is
$Nk$ on a square lattice. The denominator is instead the surface of the
cluster, i.e., the total number of ways in which a random walker can
step out of the cluster in a single step. 
We argue that the line of size $N$ is indeed the cluster configuration
with minimal exit time among all those having N nodes. In fact, any
other configuration has a smaller surface and will intuitively leave
fewer ways for the random walker to exit the cluster. Note that
Eq.~(\ref{eq:MF_exittime}) is exact on an infinite line cluster and
gives the correct value which is equal to $2$ when $N \to \infty$, as obtained in Ref.~\cite{Bassolas2021bis}. In fact, since for a line of length $N$ we have that $e=N-1$ ~\cite{Bender2010}, the surface
has to be $\sigma=2N+2$. Therefore, for
large $N$, we get: 
\begin{equation} \begin{aligned}
\lim_{N \to \infty}\tau_{MF}^{lines}=\frac{Nk}{\sigma}=\frac{4N}{2N+2}=2
\end{aligned} \label{eq:limit_lines} \end{equation}

Characterising clusters with maximal exit time at each fixed $N$ seems
to be a harder quest. Here we conjecture that the maximal exit time is
obtained for squares with $N$ nodes.  Our first assumption is that, at
fixed $N$, the cluster configuration with the maximal exit time is among
the ones with the minimal surface and tree-likeness, as shown for the
simple case of $N=4$ in Fig.~\ref{fig_N4}. We argue that the cluster
with minimal surface among the configurations with a certain $N$ is the
one that has the maximum possible number of internal $4$-cycles, where a
$4$-cycle is a sequence of $4$ distinct adjacent edges that starts and
ends at the same node. We only consider simple $4$-cycles, thus ignoring
orientation and starting nodes. A few examples of cluster configurations
with the associated $4$-cycles are shown in
Fig.~\ref{fig_squareslines}(b),(d). Note that the line is indeed one of the
configurations without $4$-cycles.

We conjecture that, in the case of a square lattice with $k=4$, given a
cluster with size $N$, if the number of its simple $4$-cycles is equal
to:
\begin{equation} \begin{aligned} \circlearrowright_{4} =
\lfloor(\sqrt{N} - 1)^2 \rfloor \end{aligned}
\label{eq:max_number_cycles} \end{equation}
then the cluster has the minimal surface for that size (see
Fig.~\ref{fig_squareslines}(b)), and if it is also a square, it also has
maximal exit time.  Notice that square clusters with $N$ (see
Fig.~\ref{fig_squareslines}(c)) trivially have minimal surface, and a
number of simple $4$-cycles equal to:
\begin{equation} \begin{aligned} \circlearrowright_{4} = (\sqrt{N} - 1)
\times (\sqrt{N} - 1) \end{aligned} \label{eq:square_number_cycles}
\end{equation}

In Fig.~\ref{fig_squareslines}(e) we show the value of
$\tau_{\mathcal{C}}$ versus the value of $\circlearrowright_{4}$ for
$10^4$ \modegm realisations with size $N=225$. The red diamond in the
top-right corner of the figure is the value of $\tau_{\mathcal{C}}$ for
a square of the same size. It is interesting to note that the square has
the highest value of $\circlearrowright_{4}$ and $\tau_{\mathcal{C}}$
over all the data points, respectively $\circlearrowright_{4} = 196$ and
$\tau_{\mathcal{C}} = 40.43$.

To reinforce this conjecture, we now show that in the limit of
$N\rightarrow\infty$, the tree-likeness of a square cluster tends to
$0.5$, which we already showed to be the lowest allowed on the square
lattice.
In Fig.~\ref{fig_squareslines}(f) we show that for a $square$ with size
$N$: (i) each of the four yellow nodes contributes with 2 edges; (ii)
each of the $\sqrt{N}-2$ nodes on the four blue stripes contributes with
3 edges; (iii) each of the  $\sqrt{N}-2 \times \sqrt{N}-2$ nodes in the
red square contributes with 4 edges. So, in the end, we have that the
number $e$ of edges in the $square$ equals:
\begin{equation} \begin{aligned} e = \frac{2\times4 + 4 \times 3 \times
(\sqrt{N}-2) + 4 \times (\sqrt{N}-2) \times (\sqrt{N}-2) }{2} = \\ = 4 +
6(\sqrt{N}-2) + 2 (\sqrt{N}-2)^2 \label{eq:square_edges} \end{aligned}
\end{equation}
If we plug this result in the formula of the tree-likeness, and we study
the limit for large $N$, we obtain:
\begin{equation} \begin{aligned} \alpha = \lim_{N\rightarrow\infty}
\frac{N-1}{ 4 + 6(\sqrt{N}-2) + 2 (\sqrt{N}-2)^2} = \frac{1}{2}
\label{eq:square_treelikeness} \end{aligned} \end{equation}
This means that, for large $N$, the $squares$ are cluster configurations
with the tree-likeness that differs the most from the value $1$, which is
associated with cluster configurations that we have proven are the ones
with the smallest value of mean exit time at fixed $N$. Furthermore,
squares are also characterised by being dense, i.e.  they are associated
with the highest value of the shape factor for that $N$. 
In the end, we have given some evidence of how the mean exit time is
related to surface, tree-likeness, and shape factor, and what set of
cluster configurations is related to minimal and maximal values of these
measures.

\subsection{Mean-Field approximation of exit times.} 

We now show that Eq.~\ref{eq:MF_exittime} provides a suitable
lower-bound estimate of the actual exit time from a generic cluster
configuration at fixed $N$.  When $N$ is small, and all the possible
configurations are known, we could in principle solve the corresponding
systems of equations to obtain the exit times.  But as $N$ increases, so
does the number of possible cluster configurations. Today, we only know
all the possible configurations of clusters for size
$N=50$~\cite{Mason2023}, so the use of numerical approximations when $N$
becomes larger than $50$, such as in our study, is necessary.
To estimate the exit time $\tau_{i,f_{\mathcal{C}}}$ from the node $i$
to the frontier $f_{\mathcal{C}}$ of a cluster realisation $\mathcal{C}$
of size $N$, we run $10^5$ random walks from each node of the cluster.
Then, we obtain the mean exit time $\tau_{f_{\mathcal{C}}}$ as an
average over the number of nodes in the cluster.
For our study, we also average the mean exit time from a cluster of size
$N$, across all the realisations of the same size, and we call this
ensemble average $\overline{\tau}$. The value of $\overline{\tau}$ is
obviously biased by the choice of the set of cluster realisations at
fixed $N$.  For small $N$ we can easily obtain and list all the possible
cluster configurations, and the $\overline{\tau}$ will reflect the
average over all the possible configurations at that size. But for large
$N$, we can only average over a certain fraction of all the possible
configurations, so the average $\overline{\tau}$ will be biased by this
sampling. In particular, in the case of \modrgm and
\modegm clusters $\overline{\tau}$ is necessarily averaged over
the configurations of clusters at fixed $N$ that the two growth models
produce with higher probability in $r$ realisations, and for this
reason, is also an indirect measure of the most probable configurations
of these growth models. We also obtain an ensemble mean-field average
exit time $\overline{\tau}_{MF}$, computed as the arithmetic mean of
Eq.~\ref{eq:MF_exittime} over all the realisations.

We generated $10^5$ cluster realisations of \modegm and \modrgm (with
$C=2$) for each $N$ in the range $[1,100]$, and we computed the
corresponding values of $\overline{\tau}$ and $\overline{\tau_{MF}}$ at
each size. We also generate $squares$ and $lines$ clusters, obtaining
their mean exit time (respectively $\overline{\tau_Q}$ and
$\overline{\tau_L}$) for each $N$.

\begin{figure}[!tbp] \centering
\includegraphics[width=1\textwidth]{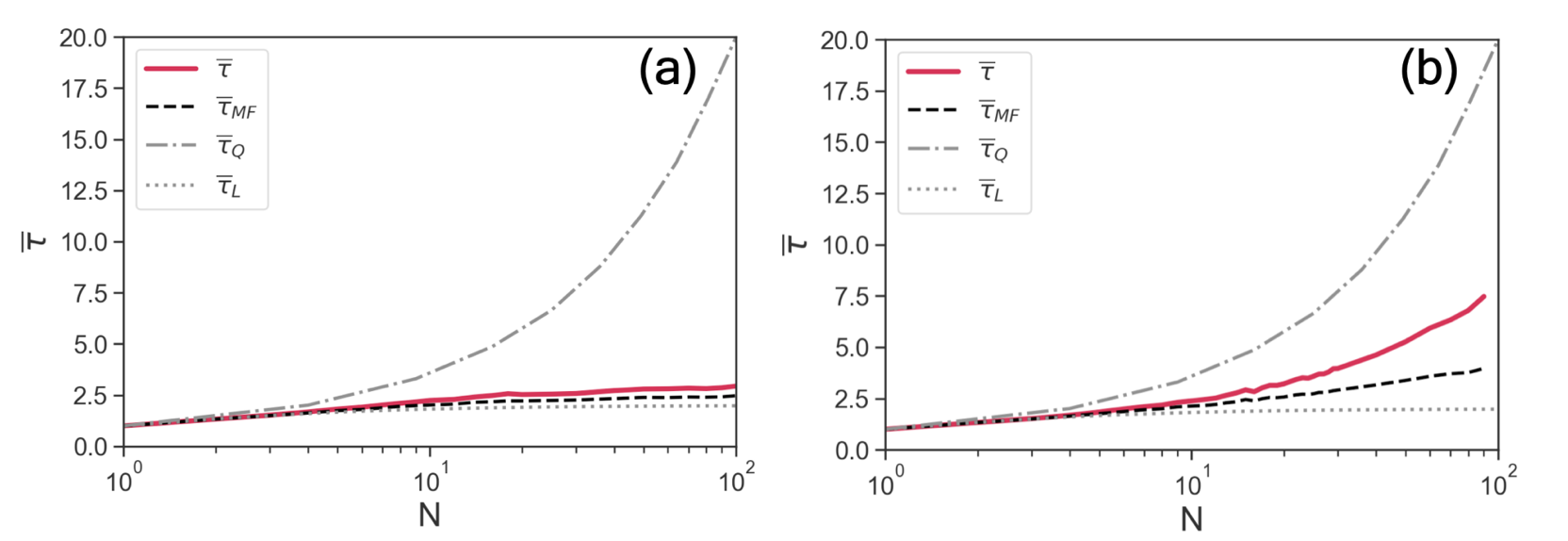}
\caption{Comparison of mean exit time for different clusters (red lines)
with that of lines $\overline{\tau}_{L}$ (dotted grey) and squares
$\overline{\tau}_{Q}$ (dash-and-dot grey) of the same size, and the
corresponding mean-field estimates $\overline{\tau}_{MF}$ based only or
cluster surface (dashed black), respectively for (a) \modrgm and (b)
\modegm (b) clusters. Note that the mean-field estimate works very well
for \modrgm clusters, which are very stripy, and still reasonably well
for small \modegm clusters.  For each size $N$, we collected $10^5$
cluster realisations, and on each realisation we computed the mean exit
time by simulating $10^5$ random walks starting from each of the nodes
in the cluster. } \label{fig_meanexit} \end{figure}

The results are shown in Fig.~\ref{fig_meanexit}, respectively for the
\modrgm (Fig.~\ref{fig_meanexit}(a)) and the  \modegm clusters
(Fig.~\ref{fig_meanexit}(b)). As expected, $\overline{\tau}$ is always
bounded by the mean exit times corresponding to the clusters of the
minimal and maximal surface. It is also evident that for \modrgm
clusters the value of $\overline{\tau}$ is quite similar to that of
lines, since the \modrgm clusters are basically stripy. The mean-field
surface approximation is less precise in the case of \modegm. This
deviation is indeed expected and due to the shape of the \modegm
clusters, which become denser and denser as $N$ increases, eventually
approaching the circle-limit profile.  Notice that $\overline{\tau}$ can
capture the structural differences between the two models: in fact, the
\modrgm mean exit time is considerably smaller than the one observed in
an \modegm cluster of the same size, as the two ensembles tend to become
more and more structurally different as $N$ increases.

\section{Mean-field theory for the size and frontier of RGM clusters}
\begin{figure}[!tbp] \centering
\includegraphics[width=1\textwidth]{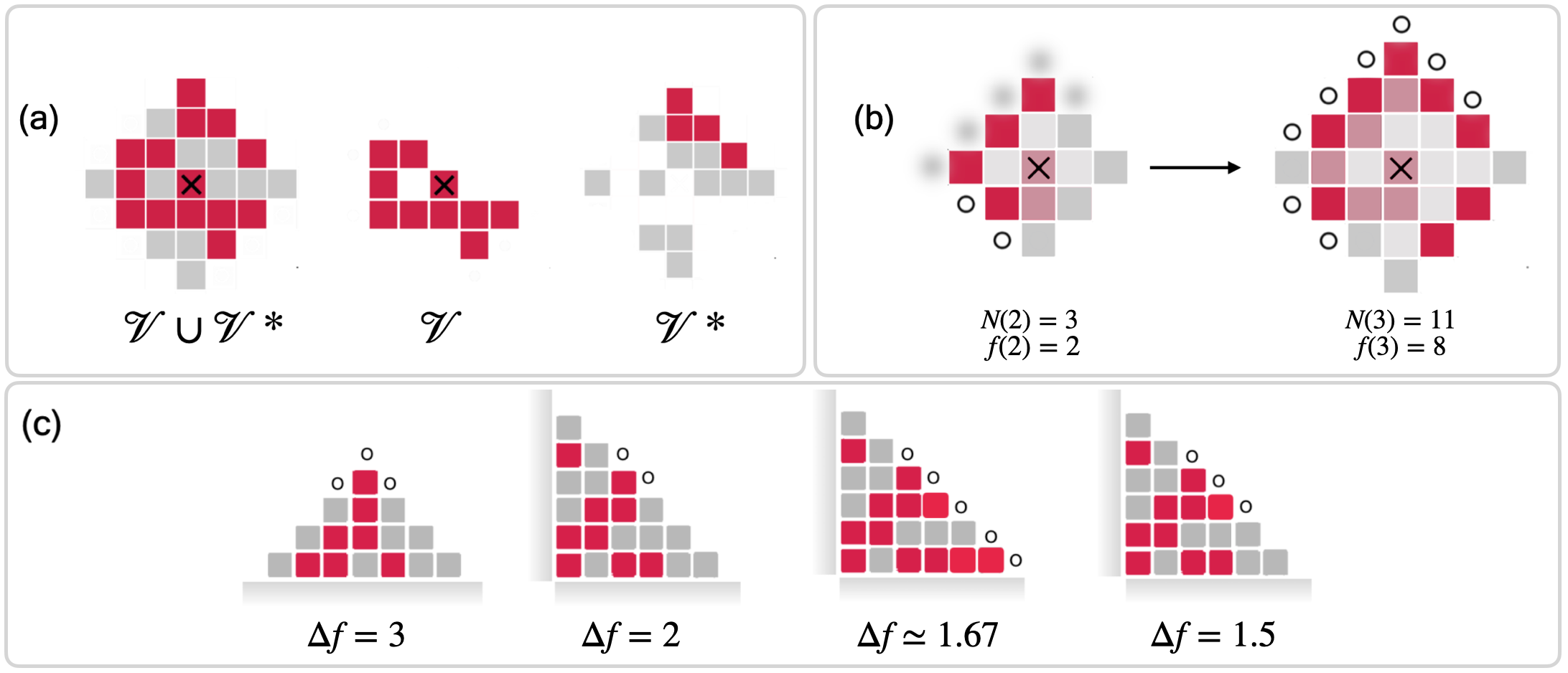} \caption{ (a) An
example of \modrgm growth after $t=3$ steps (leftmost part of the
panel).  For the sake of simplicity, neither blank nodes nor active
frontier nodes are represented in the figure. In the middle of the
panel, the seed cluster $\mathcal{V}$ marked with "$\times$" is shown.
On the rightmost part of the panel we report the coloured nodes on the
lattice that are not part of the seed cluster, namely the set
$\mathcal{V*}$.  (b) Two subsequent steps of a typical \modrgm
realisation, respectively for $t=2$ (left) and $t=3$ (right). For the
sake of simplicity, the blank nodes are not shown in the figure.  The
lighter tone represents the \modrgm cluster at the previous step.  For
both steps, $N$ and $f$ are shown. Nodes in $\mathcal{V*}$ that can
potentially become part of $\mathcal{V}$ at the next step are coloured
in blurred grey. (c) Different values of $\Delta f$ are associated with
different sets of nodes in the seed cluster active frontier. In this
picture, the shadowed stripes in each of the cases represent the bulk of
$\mathcal{V} \cup \mathcal{V*}$.  The highest value of  $\Delta f$ is
associated with the leftmost configuration. }
\label{fig_notetrends} \end{figure}

We formulate here in more detail the \modrgm growth.
Consider an infinite and blank square lattice $\mathcal{G}$, where we
assign the colour $c$ to a node chosen at random. The chosen node
becomes the so-called seed cluster, namely $\mathcal{V}$, where the
growth process begins. We denote by $N(t)$ the expected size of the
cluster $\mathcal{V}$ at the growth step $t$, and with $f(t)$ the number
of blank nodes that share at least one edge with $\mathcal{V}$ at the
same $t$, namely the $active$ $frontier$ of the cluster. We label these
special blank nodes with a "$\circ$".
When $t=1$, to each node in the active frontier of $\mathcal{V}$ is
either assigned the colour $c$ or one of the other available colours in
$C$, following the probability distribution \probP. 
If a node is coloured $c$ during the assignment, then it becomes part of
$\mathcal{V}$. We call $\mathcal{V*}$ the  union of $\mathcal{V}$ and
the other nodes coloured during the process (see
Fig.~\ref{fig_notetrends}(a)).
We now repeat the process: for each step $t\geq1$, each blank node that
shares edges with $\mathcal{V*}$ is either assigned with the colour $c$
or one of the other available colours in $C$, following the probability
distribution \probP. If a node is coloured $c$ during the assignment,
and it is adjacent to $\mathcal{V}$, then becomes part of it.
During the assignments, there is the possibility to connect
$\mathcal{V}$ with other clusters coloured with $c$ in $\mathcal{V*}$
(see Fig.~\ref{fig_notetrends}(b)). Consequently, the active frontier
$f(t)$ has to be the sum of two contributions: one from the
$\mathcal{V}$ active frontier and one from the active frontier of these
clusters in $\mathcal{V*}$.
We can write the following coupled mean-field equation for the expected
size $N(t)$ and active frontier $f(t)$ of \modrgm clusters after $t$
time steps:
\begin{equation} \left\{ \begin{aligned} N(t+1) &= N(t) +  N_e(t) + p(c)
f(t) \\ f(t+1) &= p(c) [ f(t) + f_e(t) ] \Delta f (t+1)
\end{aligned}\right.  \label{eq:mean_field} \end{equation}
The first equation states that the number of nodes $N(t)$ in
$\mathcal{V}$ is equal to the number of nodes in $\mathcal{V}$ at the
previous step, plus a number $N_e(t)$ of nodes from $\mathcal{V*}$, and
a fraction $p(c)$ of sites of $f(t)$, i.e. the ones in the active
frontier of $\mathcal{V}$ that are assigned with the $c$ colour from the
previous step.
The second equation states that the value of the active frontier
$f(t+1)$ is proportional to a fraction $p(c)$ of nodes from two
contributions: one comes from the active frontier of $\mathcal{V}$,
$f(t)$, and one is from $f_e(t)$, i.e. the active frontier of other
clusters in $\mathcal{V*}$ that can become part of $\mathcal{V}$ in the
next step. 

We multiplied these contributions by $\Delta f(t+1)$, which is the
expected increase in the size of $f(t+1)$ after the assignment of
colours at $t$. $\Delta f(t)$ is defined as the ratio between the number
of nodes in $\mathcal{V}$ sharing edges with the  frontier at time $t$
and the number of nodes in $f(t)$: it describes the average potential
increase in size due to each node in $\mathcal{V}$ that borders with
the active frontier, from one growth step to the next (see
Fig.~\ref{fig_notetrends}(c)).
We are now interested in solving these equations.
For $t=0$ we have:
\begin{align*} f(1) = p(c) [ f(0) + f_e(0) ] \Delta f(1) \end{align*}
And for $t=1$:
\begin{align*} f(2) = p(c) [ f(1) + f_e(1) ] \Delta f(2) =\notag\\ =
p(c)^2 [ f(0) + f_e(0) ] \Delta f(1) \Delta f(2) + p(c) f_e(1) \Delta
f(2) =\notag\\ = p(c)^2 f(0) \Delta f(1) \Delta f(2) + p(c)^2 f_e(0)
\Delta f(1) \Delta f(2) + p(c) f_e(1) \Delta f(2) \end{align*}

So, solving the Eq.~\ref{eq:mean_field} for $f(t)$, we obtain:

\begin{equation}
  f(t) = f(0) p(c)^t \prod_{j=1}^{t}\Delta f(j) +
  \sum_{j=0}^{t-1}\left[ p(c)^{t-j} f_e(j) \prod_{l=j+1}^{t} \Delta
    f(l) \right]
  \label{eq:solved_ft}
\end{equation}
Similarly, we can solve Eq.~\ref{eq:mean_field} for $N(t)$. For $t=0$ we have:
\begin{align*}
  N(1) = N(0) + f(0) p(c) +  N_e(0)
\end{align*}
Then, for $t=1$:
\begin{align*}
  N(2) = N(1) + f(1) p(c) +  N_e(1) =\notag\\
  = N(0) + p(c)f(0) +  N_e(0) + p(c)f(1) +  N_e(1) =\notag\\
  = N(0) + N_e(0) + N_e(1) + p(c) [ f(0) + f(1) ]
\end{align*}
And for $N(t)$:
\begin{equation}
  N(t) = N(0) + \sum_{k=0}^{t}N_e(k) + p(c) \sum_{k=0}^{t} \left[ f(0)
    p(c)^k \prod_{j=1}^{k}\Delta f(j) + \sum_{j=0}^{k-1}\left(
    p(c)^{k-j} f_e(j) \prod_{l=j+1}^{k} \Delta f(l) \right) \right]
  \label{eq:solved_nt}
\end{equation}
where we have used Eq.~\ref{eq:solved_ft} for $f(t)$. 

\begin{figure}[!tbp] \centering
\includegraphics[width=1\textwidth]{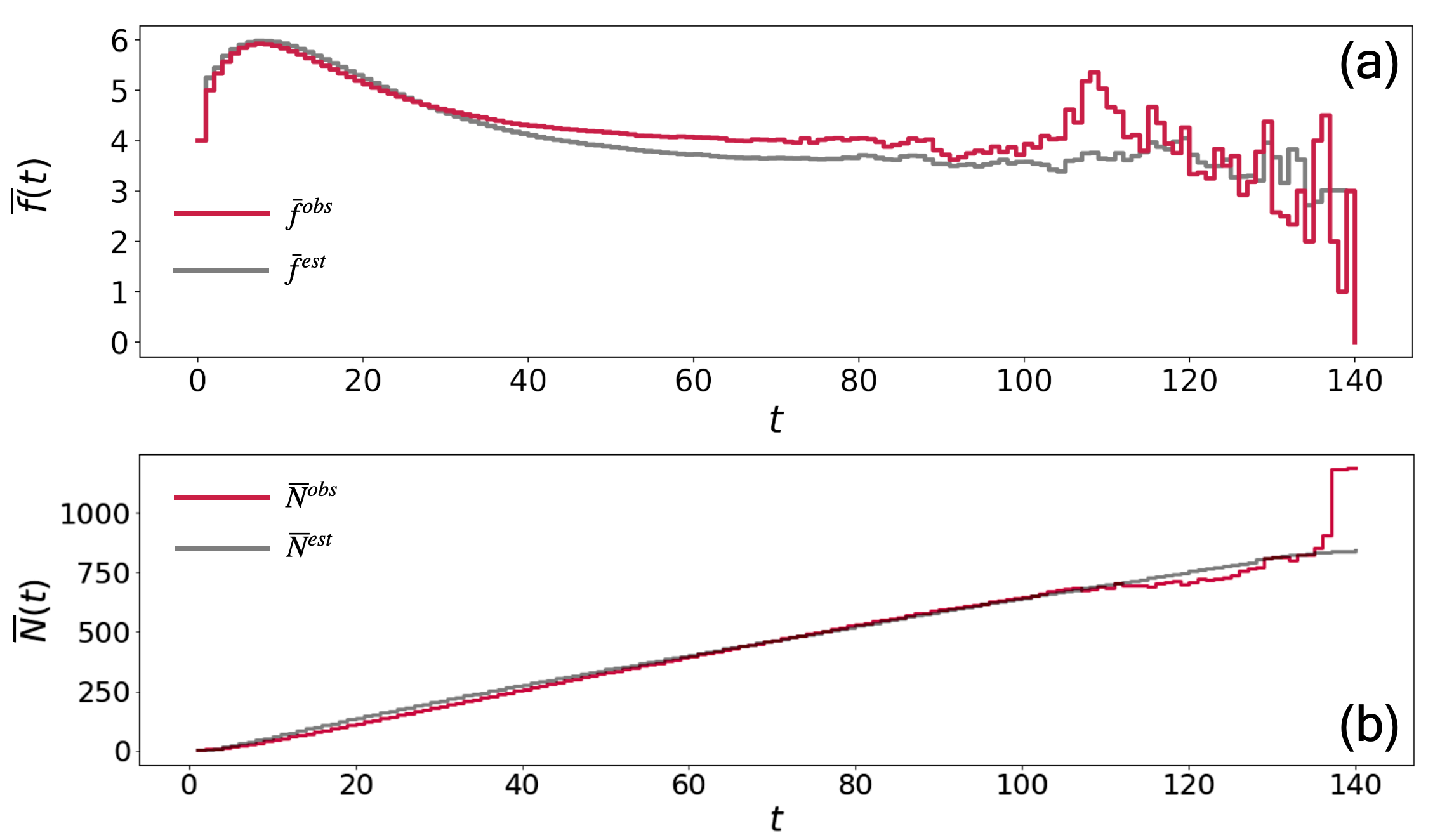} \caption{ The
observed active frontier $\overline{f}(t)^{obs}$ (a) and the observed
size $\overline{N}(t)^{obs}$ (b) of \modrgm clusters are depicted with
red solid lines. Each line is obtained from $5 \times 10^6$ realisation.
Estimated active frontier $\overline{f}(t)^{est}$ and estimated size
$\overline{N}(t)^{est}$ are shown as dark grey solid lines. Notice the
excellent agreement between the mean-field theory and the data.}
\label{fig_trends} \end{figure}

We will now propose a mean-field approach for approximating the
quantities $\Delta f(t)$ and $f_e(t)$ and obtaining an estimate for the
observed $f(t)$ and $N(t)$.
We performed $5 \times 10^6$ simulations of \modrgm  in the case
$|C|=2$. We set the maximal growth steps, i.e. the growth step at which
the process stops even if the cluster still has an active frontier, at
$t=200$. We collect all the values of $\Delta f(t)$, $f_e(t)$ and
$N_e(t)$ at each growth step $t$, and then we obtained $\overline{\Delta
f}(t)$, $\overline{f_e}(t)$ and $\overline{N_e}(t)$ as the average over
the number of different values of $\Delta f$, $f_e$ and $N_e$ at growth
step $t$.
We plug $\overline{\Delta f}(t)$ and $\overline{f_e}(t)$ in
Eq.~\ref{eq:solved_ft} to obtain our mean-field estimate for $f(t)$:
\begin{equation} \overline{f}(t)^{est} = f(0) p(c)^t \prod_{j=1}^{t}
\overline{\Delta f}(j) + \sum_{j=1}^{t-1}\left[ p(c)^{t-j}
\overline{f_e}(j) \prod_{l=j+1}^{t} \overline{\Delta f}(l) \right]
\label{eq:solved_ft_MF} \end{equation}
Notice that in Eq.~\ref{eq:solved_ft}, the sum starts from $j=0$, while
in Eq.~\ref{eq:solved_ft_MF} from $j=1$. This is because, at each time
step, the measured value of $\overline{f_e}(t)$ is nothing less than the
fraction $p(c)$ of $f_e(t)$ nodes, so the exponent of $p(c)^{t-j}$ is
reduced by one.
Then, we collect all the observed values of $f(t)$ at each growth step
$t$, and we obtain $\overline{f}(t)^{obs}$ as the averaged observed
$f(t)$ over the total number of different values of $f(t)$ at $t$. The
results are shown in Fig.~\ref{fig_trends}(a).

Note that the mean-field theory replicates quite closely the observed
trend for the active frontier, but it begins to deviate slightly after
$t=40$.  It diverges the most in the region between $t=[100,140]$,
signalling that the model somehow cannot capture the trend for large $t$,
where we have mostly fluctuations and contributions from explosive
events, i.e. $t$ steps where the cluster has grown explosively thanks to
contributions from $\mathcal{V*}$.

This discrepancy is due to a variety of factors. First, $\Delta f$ is
just an overall average value that is assumed equal for all the nodes of
the frontier. Moreover, we implicitly assumed that $\Delta f$ is the
same for nodes in $\mathcal{V}$ bordering the frontier, and nodes in
$\mathcal{V*}$ bordering blank nodes in $\mathcal{G}$: this is also
another approximation since we don't know if the two contributions are
different.
However, the mean-field theory captures very well the behaviour of
\modrgm for $t=[0,40]$. In the range $t=[40,100]$ the curve converges
towards $f(t)=4$, and the mean-field curve follows this trend quite
closely. After $t=100$, the mean-field can still provide a qualitative
estimation of the data, despite the predominance of large fluctuations.

We adopt a similar mechanism to obtain our mean-field estimate for
$N(t)$:
\begin{equation} \overline{N}(t)^{est} = N(0) +
\sum_{k=0}^{t}\overline{N_e}(k) + p(c) \sum_{k=0}^{t} \left[ f(0) p(c)^k
\prod_{j=1}^{k}\overline{\Delta f}(j) + \sum_{j=1}^{k-1}\left(
p(c)^{k-j} \overline{f_e}(j) \prod_{l=j+1}^{k} \overline{\Delta f}(l)
\right) \right] \label{eq:solved_nt_MF} \end{equation}
Also in this case, we collect all the observed values of $N(t)$ at each
growth step $t$, and we obtain $\overline{N}(t)^{obs}$ as the averaged
observed $N(t)$ over the total number of different values of $N(t)$ at
$t$. The results are shown in Fig.~\ref{fig_trends}(b).
More than in the case of $f(t)$, the model follows in a very good
agreement the data curve for the size $N(t)$ in the whole observed range
of growth steps. It deviates slightly after $t=100$, accordingly to what
we found for the tail of the active frontier, where the fluctuations become
predominant.

\subsection{Mean-field active frontier for large N.} It is easy to use a
mean-field argument to explain the observed value of the active frontier
of \modrgm clusters when $N$ becomes large. By looking at
Fig.~\ref{fig_trends}(a), we can indeed notice that in the interval
$t=[40,100]$ the  value of $\overline{f}(t)^{obs}$ is flat and can be
fitted very well by the line $\overline{f}(t)=4$. The trend diverges
from the value $4$ only for $t>100$ when the fluctuations, due to a
smaller number of points in the dataset, dominate the behaviour of the
$\overline{f}(t)^{obs}$.

\begin{figure}[!tbp] \centering
\includegraphics[width=1\textwidth]{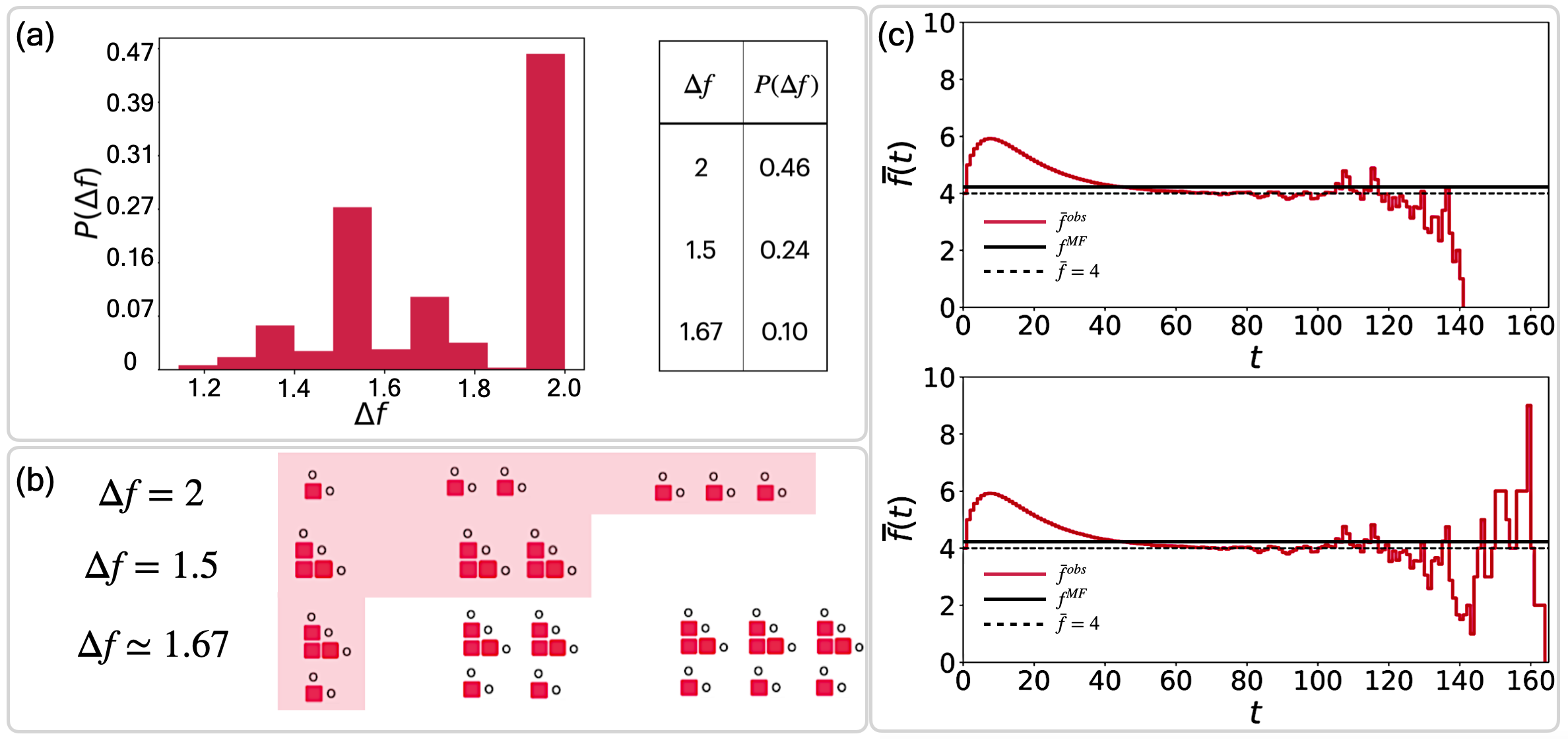} \caption{ (a) The
probability distribution for $\Delta f$ at $t=75$ after $10^7$ \modrgm
realisations. The heterogeneous distribution (left) exhibits some peaks
that represent the more probable configuration of $\Delta f$. Their
probabilities are collected in the table (right).  (b) A sketch of
\modrgm clusters, where only nodes on the border of the cluster, along
with their active frontier, are depicted. Each row is related to the
respective $\Delta f$: for example, the first row is the collection of
all the possible node-active frontier configurations that contribute to
$\Delta f = 2$. The proposed most probable configurations of active
frontier and nodes are the ones highlighted in red. (c) The
$\overline{f}(t)^{obs}$ trend for $9\times10^6$ (top) and $10^7$
(bottom) realisations of \modrgm clusters. The observed curve is in red,
while the mean-field approximation is depicted as a solid dark grey
line. $\overline{f}(t)=4$ is shown as a dashed dark grey line. }
\label{fig_ftmean} \end{figure}

In Fig.~\ref{fig_ftmean}(a) we show the probability distribution for
$\Delta f$ for the value $t=75$ after $10^7$ \modrgm realisations:
we choose this $t$ because it lies in the interval $t=[40,100]$ where
$\overline{f}(t)^{obs}$ is flat. 
The distribution appears to be extremely heterogeneous, with some peaks
that represent the more probable configuration of $\Delta f$. In the
table in Fig.~\ref{fig_ftmean}(a), we also collect the probability
associated with the three most probable $\Delta f$, along with their
value. The distribution remains very similar and stable for each $t$ in
the range $t=[40,100]$, so we make the assumption that it has to
describe the mean behaviour of $\Delta f$ when $N$ becomes large.
In Fig.~\ref{fig_ftmean}(b) we show a schematic representation of
\modrgm clusters where we only take into consideration the nodes
that are immediately adjacent to the active frontier of the cluster. We
only show the combinations of active frontiers and nodes that are
related to the three most probable $\Delta f$.
We propose that the most probable configurations of active frontier and
nodes are the ones highlighted in red in Fig.~\ref{fig_ftmean}(b),
drawing on the evidence that large \modrgm clusters are elongated
and surface-like: groups of connected nodes with a single large active
frontier are not so probable, while single or double nodes with small
active frontiers are preferred.  We use the probabilities in the table
in Fig.~\ref{fig_ftmean}(a) to weight each configuration, and we obtain
a mean-field approximation for $\overline{f}(t)^{obs}$ as:
\begin{equation} f^{MF} \simeq \frac{\frac{(2+4+6) \times 0.46}{3} +
\frac{(3+6) \times 0.24}{2} + 5 \times 0.1}{0.46+0.24+0.1} \simeq 4.27
\label{eq:frontier_MF} \end{equation}
which is in a good agreement with the value $f(t)=4$ that we observe in
the flat region of $\overline{f}(t)^{obs}$.

In figure Fig.~\ref{fig_ftmean}(c) we show the $\overline{f}(t)^{obs}$
trend for $9\times10^6$ and $10^7$  realisations of \modrgm. In both
plots, the  $\overline{f}(t)^{obs}$ overlap quite well
$\overline{f}(t)=4$ in the time range $t=[40,100]$. Furthermore, in the
bottom plot the tail of the trend presents a huge fluctuation in the
active frontier, with a maximal value of $\overline{f}^{obs}(157)=9$,
followed by an abrupt transition to zero. This finding shows that the
trend in the tail in Fig.~\ref{fig_ftmean}(c) is only ephemeral, and the
mean behaviour for the \modrgm active frontier has to be
$\overline{f}(t)=4$.

\section{Conclusions} 

Having a robust way of discriminating relevant patterns from trivial
ones is of paramount importance in any field of research. But it is
absolutely vital in the study of spatial complex systems, where the
simple existence of spatial agglomerates of similar units has been far
too often considered enough ground to conclude that some interesting
behaviour is at work. By introducing the Uniform Random Colouring (URC)
process as a random baseline for colouring lattices, we have shown that
structures with considerable size can and will emerge, even in 2D square
lattices, even in a random and uncorrelated process. This finding urges
the use of caution when measuring quantities on real coloured spatial
networks, as also very simple null models can generate large random
structures, particularly when a small number of classes is available. We
have shown that what makes clusters interesting (i.e., statistically
significant), is not just their size, but a variety of geometric
properties including their shape, tree-likeness, and surface.  The mean
exit time, i.e., the expected time needed for a uniform random walker to
escape from the cluster, seems to be a promising candidate to summarise
the geometric information about a cluster. The characterisation of
shapes with minimal and maximal exit times, that we have proposed here,
is just a first step in the exploration of this measure as an indicator
of the compactness of a cluster configuration. Indeed, a spectral
characterisation of cluster geometry might be possible, and could shed
more light on the variety of interesting patterns we observe in
real-world spatial systems. Overall, the results shown in this work
provide a solid base upon which the significance of spatial clusters can
be assessed. Given the importance of cluster analysis for a variety of
spatial problems, from segregation to resource accessibility and
distribution, the simple models and measures introduced in this work
have a potentially wide applicability.

\end{document}